\documentclass[10pt]{article}
\usepackage{fullpage}
\usepackage{amsmath}
\usepackage{amssymb}
\usepackage{amsfonts}
\usepackage{comment}
\usepackage{color}
\usepackage{cite}
\usepackage{subcaption}
\usepackage{graphicx}

\DeclareMathOperator{\sign}{sgn}

\def\##1{\underline{#1}}
\def\=#1{\underline{\underline{#1}}}

\def\+
#1{\underline{\bf #1}}
\def\*#1{\underline{\underline{\bf #1}}}

\def\r#1{(\ref{#1})}
\def\l#1{\label{#1}}
\def\c#1{\cite{#1}}

\def\le{\left(}
\def\ri{\right)}
\def\les{\left[}
\def\ris{\right]}
\def\lec{\left\{}
\def\ric{\right\}}
\def\lek{[{\kern 0.1em}}
\def\rik{{\kern 0.1em}]}

\def\.{\mbox{ \tiny{$^\bullet$} }}

\def\eps{\varepsilon}

\def\epso{\eps_{\scriptscriptstyle 0}}
\def\lambdao{\lambda_{\scriptscriptstyle 0}}
\def\muo{\mu_{\scriptscriptstyle 0}}
\def\etao{\eta_{\scriptscriptstyle 0}}

\def\ko{k_{\scriptscriptstyle 0}}

\def\ux{\hat{\#u}_x}
\def\uy{\hat{\#u}_y}
\def\uz{\hat{\#u}_z}

\def\calA{{\cal A}}
\def\calB{{\cal B}}

\def\Pmat{[{\kern 0.1em}\=P{\kern 0.1em}]}
\def\PAmat{[{\kern 0.1em}\=P_\calA{\kern 0.1em}]}
\def\PBmat{[{\kern 0.1em}\=P_\calB{\kern 0.1em}]}
\def\Pellmat{[{\kern 0.1em}\=P_\ell{\kern 0.1em}]}
\def\Ymat{[{\kern 0.1em} \=Y{\kern 0.1em}]}
\def\Imat{[{\kern 0.1em} \=I{\kern 0.1em}]}

\def\fz{[{\kern 0.1em}\#f(z){\kern 0.1em}]}

\def\epsAa{\eps_{\calA a}}
\def\epsAb{\eps_{\calA b}}
\def\epsAc{\eps_{\calA c}}
\def\epsBa{\eps_{\calB a}}
\def\epsBb{\eps_{\calB b}}

\begin{document}

\begin{center}

\Large{ {\bf  From unexceptional to doubly exceptional surface waves
}}
\end{center}
\begin{center}
\vspace{5mm} \large

  \vspace{3mm}
  
 \textbf{Akhlesh  Lakhtakia}\footnote{E--mail: akhlesh@psu.edu}\\
 {\em NanoMM~---~Nanoengineered Metamaterials Group\\ Department of Engineering Science and Mechanics\\
Pennsylvania State University, University Park, PA 16802--6812, USA}\vspace{3mm}\\

 \textbf{Tom G. Mackay}\footnote{E--mail: T.Mackay@ed.ac.uk}\\
{\em School of Mathematics and
   Maxwell Institute for Mathematical Sciences\\
University of Edinburgh, Edinburgh EH9 3FD, UK}\\
and\\
 {\em NanoMM~---~Nanoengineered Metamaterials Group\\ Department of Engineering Science and Mechanics\\
Pennsylvania State University, University Park, PA 16802--6812,
USA}\vspace{3mm}\\

\normalsize

\end{center}

\begin{center}
\vspace{15mm} {\bf Abstract}
\end{center}

An
exceptional surface wave can propagate in an isolated direction, when
guided by the planar interface of two 
homogeneous  dielectric
partnering mediums of which at least one is anisotropic,
provided
that the
 constitutive parameters of the partnering mediums satisfy certain constraints.
Exceptional surface waves are distinguished from  unexceptional surface waves by their localization characteristics: 
the fields
of an  exceptional
surface wave 
in the anisotropic partnering medium decay
as a combined linear-exponential function of distance from the interface, whereas the decay is purely
exponential for an unexceptional surface wave.
 If both partnering mediums are anisotropic then a doubly exceptional surface wave can exist  for an isolated propagation direction. The decay of  this wave in both partnering mediums
 is governed by a combined linear-exponential function of distance from the interface.

\section{Introduction}

The time-harmonic Maxwell equations for a monochromatic electromagnetic field with prescribed exponential variations in two
mutually orthogonal directions in a linear homogeneous medium can always be formulated as a 
4$\times$4 matrix ordinary differential equation \cite{Billard,Berreman,TMMEO}. Let the unit vectors
$\ux$ and $\uy$ of a Cartesian coordinate system $\#r\equiv (x,y,z)$
 be parallel to the two mutually orthogonal directions. Then, the electric and magnetic field
phasors can be expressed as 
\begin{equation} 
\label{planewave}
\left.\begin{array}{l}
 \#E (\#r)=  \#e(z)  \, 
  \exp\les i q \le x \cos \psi + y \sin \psi \ri \ris \\[4pt]
 \#H  (\#r)= \#h(z) \,
 \exp\les i q \le x \cos \psi + y \sin \psi \ri \ris 
 \end{array}\right\}\,,   
\end{equation}
where  ${q}$ is the   wavenumber in the $xy$ plane, the angle $\psi\in\left[0,2\pi\right)$,
and an $\exp(-i\omega t)$ dependence on time $t$ is implicit with $\omega$ as the angular
frequency and $i = \sqrt{-1}$.
Substitution of  the phasor representations~\r{planewave} in
the source-free Maxwell curl equations
yields
 the 4$\times$4 matrix ordinary differential
equation  
\begin{equation}
\label{MODE}
\frac{d}{dz}\fz= i \Pmat\.\fz\,,
\end{equation}
where  the   column 4-vector
\begin{equation}
\fz= 
[ 
\begin{array}{c}%
\ux\.\#e(z), \quad
\uy\.\#e(z),\quad
\ux\.\#h(z),\quad
\uy\.\#h(z)
\end{array} 
]^T\,;
\label{f-def}
\end{equation}
 the 4$\times$4 matrix
$\Pmat$ depends on $q$, $\psi$, and the constitutive parameters
of the medium; and the superscript $T$ denotes the transpose.
Both $\uz\.\#e(z)$ and $\uz\.\#h(z)$ are
 algebraically connected to $[\#f (z)]$  \c{TMMEO}.
 
Ordinarily, the matrix $\Pmat$ has four distinct eigenvalues and an eigenvector
can be prescribed for each eigenvalue such that the four eigenvectors are mutually
orthogonal. Each eigenvalue then has an algebraic multiplicity $1$ and geometric
multiplicity $1$. In some instances, 
two eigenvalues may be identical but both corresponding eigenvectors are
mutually orthogonal.
 The matrix $\Pmat$ is then
said to exhibit semisimple degeneracy, the degenerate eigenvalue having  algebraic
multiplicity $2$ and geometric multiplicity $2$ \cite{Pease,Shuvalov}. Semisimple
degeneracy is exhibited for every $\psi\in\left[0,2\pi\right)$
by the matrix $\Pmat$ formulated for free space as well as for any isotropic dielectric-magnetic
material \cite{EAB}.

In certain biaxial absorbing dielectric mediums,  $\Pmat$ may exhibit non-semisimple degeneracy
for isolated values of $\psi$, depending on the orientation of the
$x$ and $y$ axes in space \cite{Borzdov,Gerardin,Grundmann}. Then, it has only two distinct eigenvalues, each of 
algebraic
multiplicity $2$ but geometric multiplicity $1$. The observable consequences of this
non-semisimple degeneracy were experimentally demonstrated by Voigt in 1902 \cite{Voigt}
and theoretically explained by Pancharatnam in 1958 \cite{Panch1958}. A plane wave characterized
by a non-semisimple degeneracy of $\Pmat$ is called a Voigt wave \cite{Brenier,VPock}, its occurrence
showing up in the band diagram as an exceptional point \cite{Moiseyev,Heiss}. These exceptional points
can arise only if the medium of propagation is either dissipative or active \cite{Moiseyev_PRA,ML_EPJ}.

Equations~\r{planewave} and \r{MODE} are also
useful for surface-wave propagation guided by the planar interface of two homogeneous mediums
\cite{ESW_book}. Suppose that medium $\calA$  fills the half-space $z>0$ and 
medium $\calB$   the half-space $z<0$. Equation~\r{planewave} holds for all $z\in(-\infty,\infty)$
with ${q}$ as the surface  wavenumber and
the direction of propagation relative to  the $x$ axis in the $xy$ plane being prescribed by  
$\psi$. The source-free Maxwell curl equations now yield
 the 4$\times$4 matrix ordinary differential
equations \c{Berreman,TMMEO}
\begin{equation}
\label{MODE_A}
\frac{d}{dz}\fz= \left\{
\begin{array}{l}
i \PAmat\.\fz\,,  \qquad   z>0 \vspace{8pt} \\
i \PBmat\.\fz\,,  \qquad   z<0
\end{array} \,. \right.
\end{equation}
Solutions of Eqs.~\r{MODE_A} must be accepted such that the electric and magnetic fields
decay as $z\to\pm\infty$.
Additionally, the boundary condition
\begin{equation}
\label{bc}
[{\kern 0.1em}f(0^-){\kern 0.1em}]=[{\kern 0.1em}f(0^+){\kern 0.1em}]
\end{equation}
must be satisfied by the acceptable solutions of Eqs.~\r{MODE_A}.

Suppose that a  surface wave can be excited
 for a certain value of $\psi$. Then, the following four cases  
arise:
\begin{itemize}
\item Case~I: Neither $\PAmat$  has a non-semisimply degenerate eigenvalue
indicating decay
as $z\to\infty$ nor $\PBmat$ has a non-semisimply degenerate eigenvalue
indicating decay
as $z\to-\infty$;
\item Case~II: $\PAmat$ has a non-semisimply degenerate eigenvalue indicating decay as $z\to\infty$,
but $\PBmat$ does not have {non-semisimply} degenerate eigenvalues indicating decay as $z\to-\infty$;
\item Case~III: $\PAmat$ does not have {non-semisimply} degenerate eigenvalues indicating decay as $z\to\infty$,
but $\PBmat$
has a non-semisimply degenerate eigenvalue indicating decay as $z\to-\infty$;
\item Case~IV: $\PAmat$ has a non-semisimply degenerate eigenvalue indicating decay as $z\to\infty$
and $\PBmat$
has a non-semisimply degenerate eigenvalue indicating decay as $z\to-\infty$.
\end{itemize}
The surface wave can be classified as:
\begin{itemize}
\item \textit{unexceptional} if Case~I holds, 
\item \textit{exceptional} if either Case~II or Case~III holds,
and 
\item \textit{doubly exceptional} if Case~IV holds.
\end{itemize}

Unexceptional surface waves are commonplace in the electromagnetics literature
\cite{ESW_book,Boardman}, 
having been theoretically established  by Uller in 1903 \cite{Uller,Zenneck,Uller-Zenneck}.
Common examples are surface-plasmon-polariton (SPP)
waves \cite{Sprokel,Fantino,Zhang,Durach}
and Dyakonov surface waves  \cite{MSS,Dyakonov88,Takayama_exp,Furs}. 
These surface waves may exist   either for   every $\psi\in\left[0,2\pi\right)$ \cite{Sprokel,Fantino,Zhang,Durach}
or for restricted ranges of $\psi$ \cite{Takayama_rev,ML_IEEE-PJ}.

The concept
of exceptional surface waves is exemplified by SPP--Voigt waves
\cite{ZML_PRA} and Dyakonov--Voigt surface waves \cite{MZL_PRSA,ZML_JOSAB}.
These surface waves arise from the non-semisimple degeneracy of either $\PAmat$ or $\PBmat$
but not of both.
Notably, the existence
of exceptional surface waves can be supported by nondissipative (and inactive) mediums, unlike the Voigt waves
\cite{Borzdov,Gerardin,Grundmann,Voigt,Panch1958,Brenier,VPock}
that are their plane-wave cousins.

The novelty of this paper is the introduction of doubly exceptional surface waves, for which
both $\PAmat$  and $\PBmat$ exhibit non-semisimple degeneracy for the same
value of $\psi$. For definiteness, in Sec.~\ref{theory}   medium $\calA$ is an orthorhombic dielectric material
whereas medium $\calB$ is a uniaxial dielectric material. Numerical examples are provided
in Sec.~\ref{nrd} following by concluding remarks in Sec.~\ref{cr}. 
The permittivity and permeability of free space are written as $\epso$ and $\muo$, respectively, so that
$\ko = \omega \sqrt{\epso \muo}$ is the
 free-space wavenumber and $\etao = \sqrt{\muo/\epso}$ is the intrinsic impedance of free space.
The  operators $\mbox{Re} \lec \. \ric$ 
and $\mbox{Im} \lec \. \ric$ deliver the
real and imaginary parts, respectively, of  complex-valued quantities and
 the complex conjugate is denoted by an asterisk.

\section{Illustrative Theory}\label{theory}

Although the concept of doubly exceptional surface waves is general enough to encompass linear
bianisotropic materials \cite{EAB}, medium $\calA$ is taken
as an orthorhombic dielectric material with
relative permittivity dyadic \cite{Nye}
\begin{equation}
\label{epsA}
\={\eps}_\calA =   \epsAa \,\uz\,\uz +
  \epsAb\, \ux\,\ux +  \epsAc\, \uy\,\uy 
\,
 \end{equation}
and medium $\calB$ as a uniaxial dielectric material with relative permittivity dyadic \cite{Nye}
\begin{equation}
\label{epsB}
\={\eps}_\calB =   \epsBa \,\uz\,\uz +
  \epsBb\, \ux\,\ux +  \epsBa\, \uy\,\uy 
\,,
 \end{equation}
 for the sake of illustration.
 We take all five principal relative permittivity scalars $\epsAa$, $\epsAb$,   $\epsAc$, $\epsBa$, and $\epsBb$
 as positive real. Furthermore,  $\epsAb > \epsAa > \epsAc$ so that
 both optic ray axes of medium $\calA$   lie in the $xy$ plane with the $y$ axis as the
bisector,
the angle  $\delta_\calA =$ 
$ \cos^{-1} $ $\sqrt{\le\epsAb-\epsAa \ri/\le\epsAb - \epsAc\ri}$ being
the half-angle  between the two optic ray axes.
 The sole optic ray axis of   medium $\calB$ is parallel to the
 $x$ axis.  Parenthetically, since mediums $\calA$ and $\calB$ are both anisotropic,
  the possibility of semisimple degenerate eigenvalues does not arise for
$\PAmat$ or  $\PBmat$.

\subsection{Fields in medium $\calA$}
The  4$\times$4 propagation matrix  $\PAmat$ can be written compactly as
\begin{equation}
\PAmat = \les
\begin{array}{cc}
\lek\=0\rik & \lek\=\varrho_{\calA}\rik
\\
\lek\=\varkappa_{\calA}\rik & \lek\=0\rik
\end{array}
\ris\,,
\l{PAmat1}
\end{equation}
where $\lek\=0\rik$ is the 2$\times$2 null matrix,
\begin{equation}
\lek\=\varrho_{\calA}\rik = \frac{1}{ \omega \epso \epsAa}  \les   
\begin{array}{cc}
\displaystyle{ {q^2 \cos \psi \sin \psi} }& 
\displaystyle{  {\ko^2 \epsAa- q^2  \cos^2 \psi}   }   \\
 \displaystyle{ {q^2  \sin^2 \psi -\ko^2 \epsAa } }&
\displaystyle{-  {q^2 \cos \psi \sin \psi} }   
\end{array}
\ris\,,
\l{PAmat2}
\end{equation}
and
\begin{equation}
\lek\=\varkappa_{\calA}\rik= \frac{1}{\omega \muo}  \les   
\begin{array}{cc}
 \displaystyle{-q^2 \cos \psi \sin \psi} & 
\displaystyle{{q^2  \cos^2 \psi-\ko^2 \epsAc}}    \\
\displaystyle{{\ko^2   \epsAb- q^2  \sin^2 \psi }}  &
\displaystyle{q^2 \cos \psi \sin \psi} 
\end{array}
\ris\,.
\l{PAmat3}
\end{equation}

\subsubsection{Non-degenerate  $\PAmat$}
When $\PAmat$ is  non-degenerate  for a specific $\psi$, its four eigenvalues are
 $\alpha_{\calA 1}$, $  \alpha_{\calA 2}$,  $\alpha_{\calA 3}=- \alpha_{\calA 1}$ and $\alpha_{\calA 4}=- \alpha_{\calA 2}$ with
\begin{equation} 
\l{a_decay_const}
\left.
\begin{array}{l}
\alpha_{\calA 1} = i \sqrt{ A_1  +A_2 q^2+\sqrt{A_3 + A_4 q^2 + A_5 q^4}} \vspace{8pt}\\
\alpha_{\calA 2} = i \sqrt{ A_1  +A_2 q^2-\sqrt{A_3 + A_4 q^2 + A_5 q^4}}
\end{array}
\right\}
\end{equation}
chosen such that ${\rm Im}\lec\alpha_{\calA 1}\ric>0$ and ${\rm Im}\lec\alpha_{\calA 2}\ric>0$.
The scalar quantities
\begin{eqnarray}
&& \hspace{-5mm}
A_1 =\displaystyle{ -\le{1}/{2}\ri\ko^2\le\epsAb+\epsAc\ri} \,,
\\
&& \hspace{-5mm}
A_2 =  \displaystyle{\le{1}/{2}\ri \le 1+ \frac{\epsAb\cos^2\psi+\epsAc \sin^2 \psi}{ \epsAa}\ri} \,,
\\
&& \hspace{-5mm}
A_3= \displaystyle{ \le{1}/{4}\ri\ko^4 \le \epsAb-\epsAc \ri^2}\,,
\\
&& \hspace{-5mm}  
A_4=\displaystyle{ \ko^2\le   \epsAc-\epsAb \ri} 
 \displaystyle{  \le \frac{    \epsAb\cos^2\psi+\epsAa \sin^2 \psi - A_6
}
{2 \epsAa}\ri
} \,,
\\
&& \hspace{-5mm}
A_5 =\displaystyle{\le{1}/{4}\ri \le1-\frac{\epsAb\cos^2\psi+\epsAc   \sin^2 \psi }{ \epsAa}\ri^2}  ,
\end{eqnarray}
and
\begin{eqnarray}
&&
A_6 = \epsAa\cos^2\psi+\epsAc \sin^2 \psi
\end{eqnarray}
are independent of $q$. Linearly independent eigenvectors of $\PAmat$ are as follows:  
\begin{equation}
\lek\#v_{\calA \ell}\rik=
\les
\begin{array}{c}
\displaystyle{\frac{\ko^2  \epsAa - \alpha_{\calA \ell}^2 - q^2 }{\ko \alpha_{\calA \ell}
\le \epsAa - \epsAb \ri }} \vspace{8pt}
\\
\displaystyle{\frac{ q^2 \le \alpha^2_{\calA \ell} + q^2 -  \ko^2\epsAa     \ri  \sin \psi \cos \psi}
{\ko \alpha_{\calA \ell} \les \epsAa \le  \ko^2 \epsAc-\alpha^2_{\calA \ell} \ri - q^2  {A_6 }\ris }} \vspace{8pt} \\
\displaystyle{\frac{q^2 \le \epsAa-\epsAc \ri \sin \psi \cos \psi }{ {\etao \les \epsAa \le  \ko^2 \epsAc-\alpha^2_{\calA \ell} \ri 
- q^2  A_6  \ris }}} \vspace{8pt}\\
\displaystyle{\etao^{-1}}
\end{array}
\ris, 
\qquad \ell \in \les 1,  4 \ris.
\end{equation}
Hence, the   solution of Eq.~\r{MODE_A}${}_1$   is given as
\begin{eqnarray}&& \nonumber
 \l{D_gen_solA}
\les\#f (z)\ris =C_{\calA 1}  \lek\#v_{\calA 1}\rik \exp \le i \alpha_{\calA 1} z \ri +  
C_{\calA 2} \lek \#v_{\calA 2}\rik \exp \le i \alpha_{\calA 2} z \ri \,,\\
&& \hspace{50mm}
\quad z > 0\,,
\end{eqnarray}
for   fields that decay as $z \to +\infty$, with
 $C_{\calA 1}\in\mathbb{C}$ and $C_{\calA 2}\in\mathbb{C}$ 
as constants   to be determined using Eq.~\r{bc}.

\subsubsection{Non-semisimply degenerate $\PAmat$}
When $\PAmat$ exhibits
non-semisimple degeneracy for a specific value of $\psi$, it has only two 
distinct eigenvalues   denoted by $\pm  \alpha_\calA$,
each with algebraic multiplicity $2$ and geometric multiplicity $1$ \c{Pease}.
Furthermore, either $\alpha_\calA = s_\calA^+$ or
$\alpha_\calA = s_\calA^-$, where
 \begin{equation} 
 \l{alphaa_sol}
s^\pm_\calA = \displaystyle{i \sign({\cos\psi}) \sqrt{A_1+A_2 \le q^\pm_\calA \ri^2 }}\in\mathbb{C}\,,
\end{equation}
 \begin{equation} 
\l{qa_sol}
q^\pm_\calA = \sign({\cos\psi}) \displaystyle{\sqrt{\frac{-A_4 \pm \sqrt{A_4^2-4 A_3 A_5}}{2 A_5}}} \in\mathbb{R},
 \end{equation} 
 $\sign(\.)$ is the signum function, and the sign of the
 square root in Eq.~\r{alphaa_sol} must be chosen  so that ${\rm Im} \lec
 \alpha_\calA \ric>0$.
   Non-semisimple degeneracy cannot occur when
$\psi =\ell\pi/2$, $\ell\in\lec{0,1,2,3}\ric$. 

An eigenvector of $\PAmat$  corresponding to the eigenvalue $\alpha_\calA=s^\pm_{\calA}$ 
can be written as 
\begin{eqnarray} 
\nonumber
&& \hspace{-5mm}
\lek\#v^\pm_{\calA } \rik= 
\les \begin{array}{c}
\displaystyle{-\frac{A_7}
{\ko s^\pm_\calA\epsAa  \le \epsAa-\epsAb \ri}} \vspace{8pt}\\
\displaystyle{\frac{ \le  q_\calA^\pm \ri^2  A_7 \sin \psi\cos\psi}
{ \ko s^\pm_\calA \epsAa \lec \epsAa \les \ko^2  \epsAc -\le s^\pm_\calA \ri^2 \ris - \le q_\calA^\pm\ri^2 A_6 \ric }} \vspace{8pt}\\
\displaystyle{\pm  {\etao}^{-1} \sqrt{\frac{\epsAa-\epsAc}{\epsAb-\epsAa}}} \vspace{8pt} \\  \etao^{-1}
\end{array}
\ris
\l{vA}\\ &&
\end{eqnarray} 
and the corresponding generalized eigenvector \c{Pease,Boyce}  as  
\begin{eqnarray} \l{wA}
\nonumber
&&
\hspace{-10mm}
\lek\#w^\pm_{\calA }\rik = \\
\nonumber
&&
\hspace{-10mm}
\les \begin{array}{c}
\displaystyle{\frac{A_7 + A_8}{ \ko A_9   \le \epsAa - \epsAb \ri} }
\vspace{8pt}\\
\displaystyle{\frac{\ko^2 \le \epsAa -\epsAb\ri A_9 +  \le A_7+A_8 \ri \les \le q_\calA^\pm\ri^2 \sin^2 \psi - \ko^2 \epsAb \ris}{\ko  A_9 
 \le q_\calA^\pm\ri^2  \le \epsAa - \epsAb \ri \sin \psi \cos \psi }} \vspace{8pt}\\
\displaystyle{\frac{ \ko \epsAa \le s^\pm_{\calA } \ri^2 \le A_7 +A_8 \ri -A_7 A_9}{\etao  s^\pm_{\calA } A_9 \le q_\calA^\pm\ri^2 
  \le \epsAa - \epsAb \ri \sin \psi \cos \psi
   } }
 \vspace{8pt} \\ 0
\end{array}
\ris,\\
&&
\end{eqnarray}
where  
\begin{eqnarray}
&& \hspace{-5mm}
A_7 = \displaystyle{\epsAa \les \le q_\calA^\pm\ri^2 +  \le s^\pm_\calA \ri^2 - \ko^2 \epsAa \ris} \,,
\\
\nonumber &&
\hspace{-5mm}
A_{8} =  \displaystyle{\ko^2 \epsAc \le \epsAb-\epsAa \ri+\le q_\calA^\pm\ri^2 \cos^2 \psi  }
\\
&&
\hspace{-5mm}
\displaystyle{\times\les \epsAa-\epsAb
 \pm \sqrt{\le \epsAb-\epsAa \ri \le \epsAa- \epsAc \ri}\tan \psi\ris}\,,
\end{eqnarray}
and 
\begin{eqnarray}
&& \nonumber
A_9= \displaystyle{\le s^\pm_\calA \ri^2 \epsAa- \ko^2 \epsAb \epsAc}
\\
&&\qquad \displaystyle{
+\le q_\calA^\pm\ri^2 \le \epsAb \cos^2 \psi + \epsAc \sin^2 \psi \ri } \,.
\end{eqnarray}
Hence, the  solution of Eq.~\r{MODE_A}${}_1$ is expressed as 
\begin{eqnarray} 
\nonumber
&&  \hspace{-5mm}
\fz = \les  C_{\calA 1} \lek\#v_{\calA }^\pm\rik  + \ko C_{\calA 2}
\le i z \lek\#v_{\calA }^\pm\rik+ 
\lek\#w_{\calA}^\pm\rik \ri
\ris \exp \le i s^\pm_{\calA } z \ri\,,
\\
&&  \hspace{50mm} z > 0\,,
\l{DV_gen_solA}
\end{eqnarray}
instead of  Eq.~\r{D_gen_solA},
  for fields that decay as $z \to +\infty$, where  all the upper signs hold if $\alpha_\calA=s^+_\calA$
  and all the lower signs   if $\alpha_\calA=s^-_\calA$.

\subsection{ Fields in medium $\calB$}
The matrix $\PBmat$ can be obtained 
on replacing $\epsAa$ by $\epsBa$, $\epsAb$ by $\epsBb$,
and $\epsAc$ by $\epsBa$ in Eqs.~\r{PAmat1}--\r{PAmat3}  provided for  $\PAmat$.

\subsubsection{Non-degenerate  $\PBmat$}
When $\PBmat$ is not non-semisimply degenerate for a specific $\psi$, its four eigenvalues are
 $\alpha_{\calB 1}$, $ \alpha_{\calB 2}$,  $\alpha_{\calB 3}=- \alpha_{\calB 1}$, and $
 \alpha_{\calB 4} = - \alpha_{\calB 2}$, with
 \begin{equation}
  \l{b_decay_const}
\left.
\begin{array}{l}
\alpha_{\calB 1} = i \sqrt{ q^2 - \ko^2 \epsBa} \vspace{8pt}\\
\alpha_{\calB 2} = \displaystyle{
i\sqrt{{q^2   \le   \sin^2 \psi + \frac{\epsBb}{\epsBa}  \cos^2 \psi \ri  -  \ko^2   \epsBb}}
}
\end{array}
\right\}\,
\end{equation}
chosen such that ${\rm Im}\lec  \alpha_{\calB \ell} \ric >0$, $\ell\in\lec1,2\ric$.
Linearly independent eigenvectors of  $\PBmat$  corresponding to these   eigenvalues  
are  
\begin{eqnarray}
\nonumber
&& \vspace{-5mm}
\lek\#v_{\calB \ell}\rik = 
\les \begin{array}{c}
0 \vspace{8pt}\\
\displaystyle{
\frac{\ko \alpha_{\calB \ell}}{q^2 \sin\psi\cos\psi}} \vspace{8pt}\\
\displaystyle{\etao^{-1} \le\cot \psi - \frac{ \ko^2 \epsBa }{q^2 \sin\psi\cos\psi}\ri }\vspace{8pt}
 \\ \etao^{-1}
\end{array}
\ris\,,
\quad \ell\in\lec1,3\ric\,,
\end{eqnarray}
and
\begin{equation}
\lek\#v_{\calB \ell}\rik = 
\les \begin{array}{c}
\displaystyle{1 - \frac{q^2  \cos^2 \psi }{\ko^2\epsBa }
} \vspace{8pt}\\
\displaystyle{-\frac{q^2  \cos  \psi \sin \psi }{ \ko^2 \epsBa }} \vspace{8pt}\\ 0 \vspace{8pt} \\ 
\displaystyle{\frac{\alpha_{\calB \ell}}{\omega \muo}}
\end{array}
\ris\,,\quad \ell\in\lec2,4\ric\,.
\end{equation}
Thus, the  solution for Eq.~\r{MODE_A}${}_2$   for fields that decay as $z \to -\infty$ is given as
\begin{equation} 
\l{D_gen_solB}
\fz =C_{\calB 3}  \lek\#v_{\calB 3}\rik \exp \le i \alpha_{\calB 3} z \ri +  C_{\calB 4} 
\lek\#v_{\calB 4}\rik \exp \le i \alpha_{\calB 4} z \ri \,,\quad z < 0\,,
\end{equation}
wherein the constants $C_{\calB 3}\in\mathbb{C}$ and $C_{\calB 4}\in\mathbb{C}$ have to
be determined by using the boundary condition 
\r{bc}.

\subsubsection{Non-semisimply degenerate $\PBmat$}

When $\PBmat$ exhibits
non-semisimple degeneracy for a specific value of $\psi$, 
$\alpha_{\calB 1}=\alpha_{\calB 2}=\alpha_\calB$
and $\alpha_{\calB 3}=\alpha_{\calB 4}=-\alpha_\calB$, with
\begin{equation} 
\l{alphab_sol}
\alpha_\calB = i \sign(\cos\psi) \ko \sqrt{\epsBa} \tan \psi
\end{equation}
chosen such that ${\rm Im}\lec\alpha_\calB\ric>0$ and
\begin{equation}
\label{qb_sol}
q = \sign(\cos\psi) \frac{\ko  \sqrt{\epsBa}}{\cos \psi}\,.
 \end{equation}  
Non-semisimple degeneracy is not possible for $\psi\in\lec0,\pi\ric$.

An eigenvector of
matrix $\PBmat$  corresponding to the eigenvalue $-\alpha_{\calB}$ 
is
\begin{equation} 
\l{vB}
\lek\#v_{\calB }\rik = 
\les \begin{array}{c}
0 \vspace{8pt}\\
\displaystyle{\frac{i \sign(\cos\psi)  }{\sqrt{ \epsBa}}} \vspace{8pt}\\
0 \vspace{8pt} \\ \etao^{-1}
\end{array}
\ris;
\end{equation}
and the corresponding generalized eigenvector   \c{Boyce,Pease} is 
\begin{equation} 
\l{wB}
\lek\#w_{\calB }\rik =  \frac{2}{\ko \le \epsBb - \epsBa \ri}
\les \begin{array}{c}
\displaystyle{  1}
\vspace{8pt}\\
\displaystyle{\tan\psi \les 1 -\le 1+ \frac{\epsBb}{\epsBa}\ri \frac{\cot^2\psi}{2}\ris
} \vspace{8pt}\\
\displaystyle{  i \etao^{-1}\sign(\cos\psi) \sqrt{\epsBa}}
 \vspace{8pt} \\ 0
\end{array}
\ris.
\end{equation}
Thus, the   solution of Eq.~\r{MODE_A}$_{2}$   for fields that decay as $z \to -\infty$ is given as
\begin{eqnarray} 
\nonumber
&&  \hspace{-5mm}
\fz = \les  C_{\calB 3} \lek\#v_{\calB }\rik  + \ko C_{\calB 4}
\le i z \lek\#v_{\calB }\rik+ 
\lek\#w_{\calB}\rik \ri
\ris \exp \le - i \alpha_{\calB } z \ri\,,
\\
&&  \hspace{50mm} z < 0\,,
\l{DV_gen_solB}
\end{eqnarray}
with unknown   constants $C_{\calB 3}$ and $C_{\calB 4}$.

\subsection{ Boundary condition at the interface $z=0$}
Equation~\r{bc} delivers
\begin{equation}
\label{eq36}
\Ymat \. \les \:
 C_{\calA 1}, \quad
  C_{\calA 2}, \quad
   C_{\calB 3}, \quad
    C_{\calB 4} \:
 \ris^T =  \les \:
 0, \quad
  0, \quad
   0, \quad
    0 \:
 \ris^T,
\end{equation}
wherein
the 4$\times$4 characteristic matrix $\Ymat$ must be singular for  surface-wave propagation \c{ESW_book}. The resulting
dispersion equation 
\begin{equation}
\l{dispersion_eq}
\left\vert \Ymat\right\vert = 0
 \end{equation} 
must be satisfied for a surface wave to propagate in the direction parallel to $\ux\cos\psi+\uy\sin\psi$.
If $\psi$ is replaced by $-\psi$ or by $\pi \pm \psi$ then 
the dispersion equation~\r{dispersion_eq} is unchanged. Accordingly,   attention can be restricted to the quadrant $0 \leq \psi \leq \pi/2$. 

\section{Discussion and Numerical Illustrations} \label{nrd}
The distinction between unexceptional, exceptional, and doubly exceptional surface waves can be clarified now. 
For numerical illustrations, 
the free-space
wavelength $\lambdao=2\pi/\ko$ was fixed equal to $633$~nm.

Let us begin by choosing  medium $\calA$  to be crocoite in its orthorhombic form \c{Collotti},
so that $\epsAa=5.6169$, $\epsAb=7.0756$, and $\epsAc=5.3361$. The constitutive parameters of medium $\calB$ are given as
$\epsBa = 5.6791$ and $\epsBb = \epsBa + \delta$, with $\delta \ge 0$. Also, the propagation angle $\psi$ is fixed at $31^\circ$.

\begin{figure}[!htb]
\centering
\includegraphics[width=6.3cm]{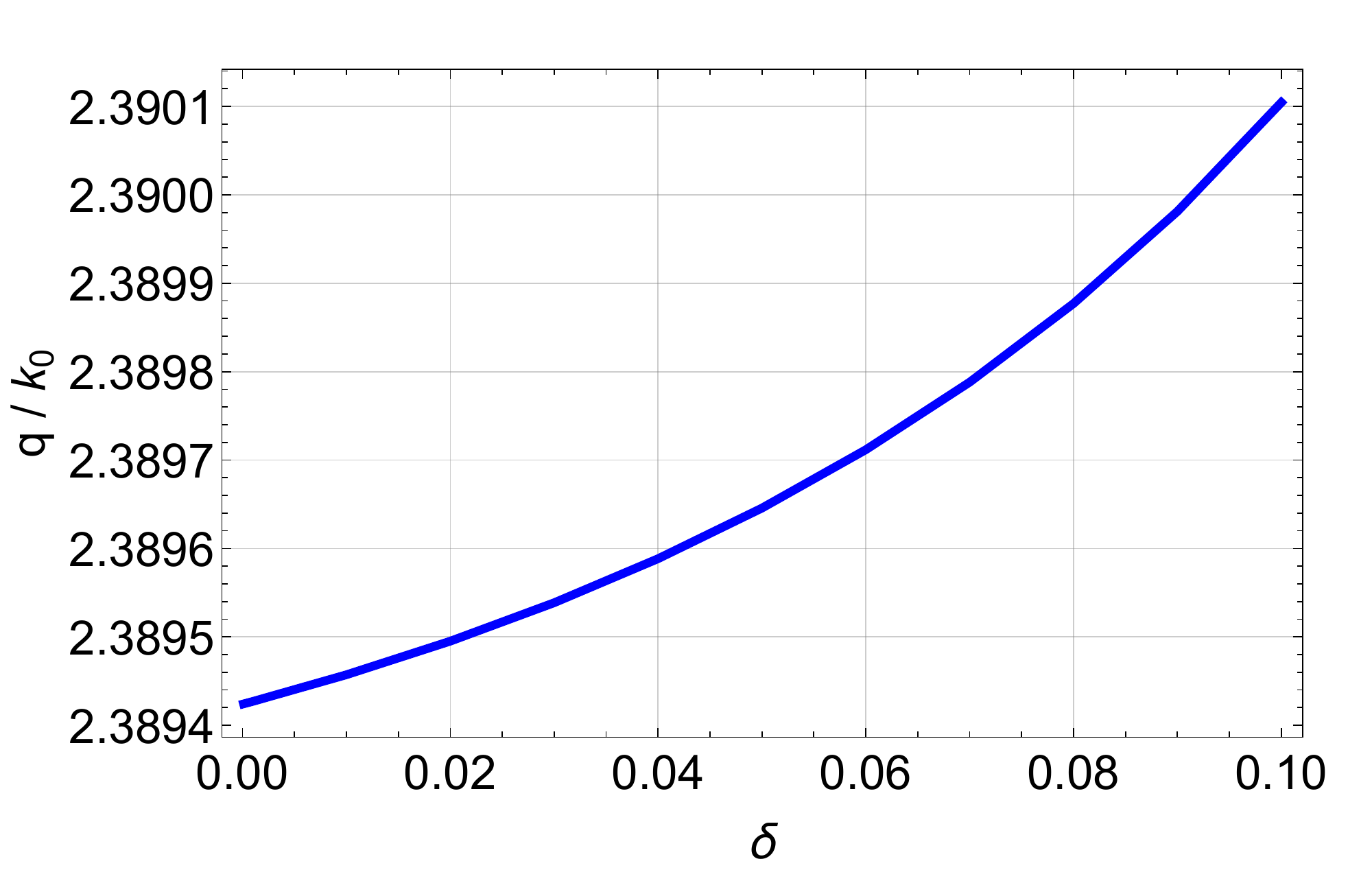}
 \caption{$q/\ko$ as a function of $\delta$ when medium $\calA$ is crocoite and medium $\calB$ is specified by $\epsBa = 5.6791$ and $\epsBb = \epsBa + \delta$. The propagation angle $\psi = 31^\circ$.} \label{fig1}
\end{figure}

For this example, 
unexceptional surface waves conforming to Case~I exist for $\delta > 0$ while an
exceptional surface wave conforming to  Case~II exists for $\delta =0$.
For the unexceptional surface waves,
the fields in the   half-spaces $z>0$ and $z<0$ are specified through  Eqs.~\r{D_gen_solA} and \r{D_gen_solB}, respectively.
 Equation~\r{dispersion_eq}
has be solved for $q$, often numerically using, for example,   the Newton--Raphson method \c{N-R}. 
For the exceptional surface wave,
the fields in the   half-spaces $z>0$ and $z<0$ are specified through  Eqs.~\r{DV_gen_solA} and \r{D_gen_solB}, respectively.
However, Eq.~\r{dispersion_eq} does not have to be solved to determine $q$; instead, $q$ is provided by
Eq.~\r{qa_sol}. In both cases,   Eq.~\r{eq36} has to be manipulated to determine
three of the four coefficients in the column 4-vector on the left side 
in terms of the fourth coefficient.

The relative surface wave number $q/\ko$ is plotted against  $\delta$ in Fig.~\ref{fig1} for $\delta \in \les 0, 0.1 \ris$. 
The value of  $q/\ko$ increases uniformly as $\delta$ increases to $0.1$. 
No surface-wave solutions were found for   $\delta>0.1$. 

\begin{figure}[!htb]
\centering
\includegraphics[width=4.3cm]{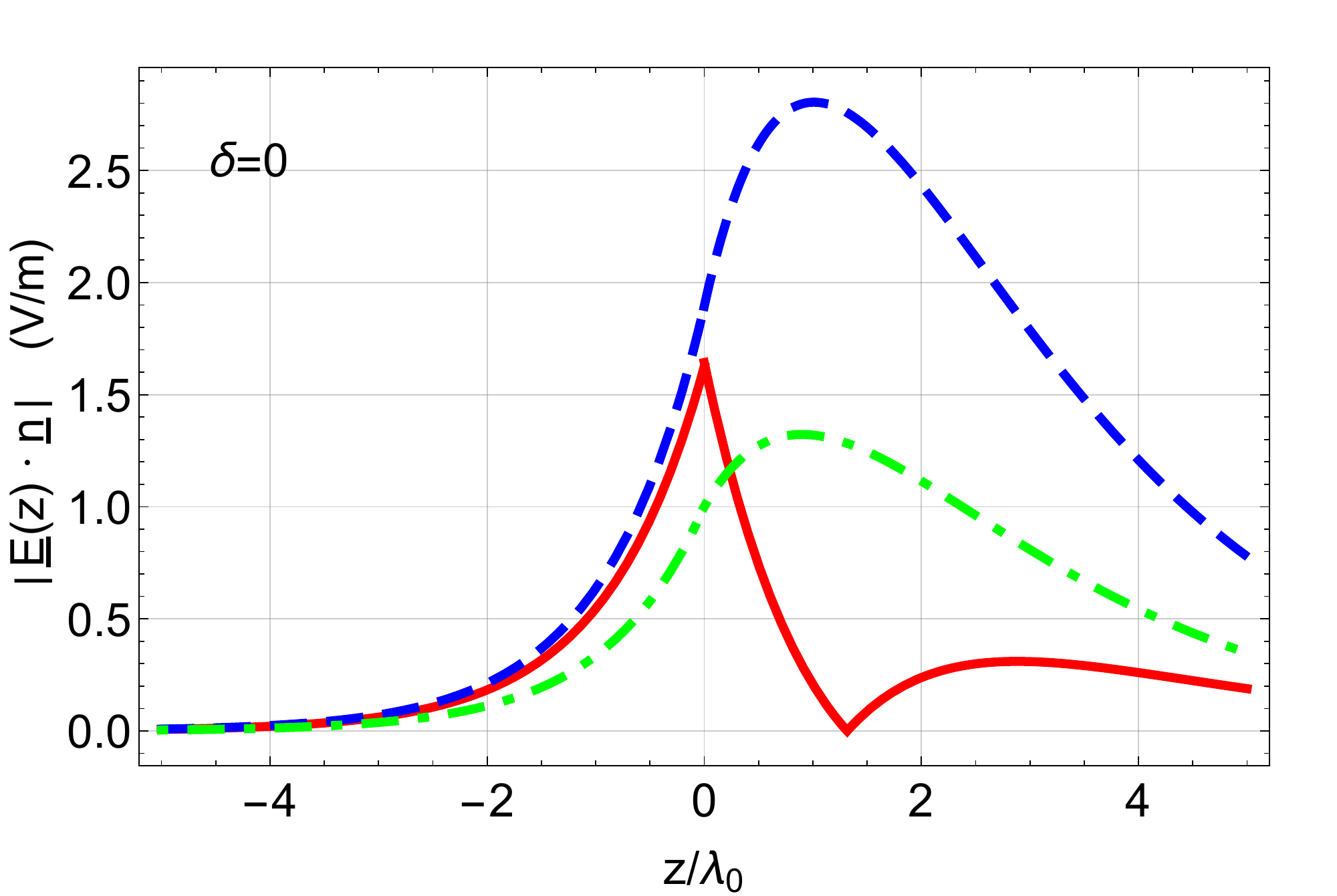}
\includegraphics[width=4.3cm]{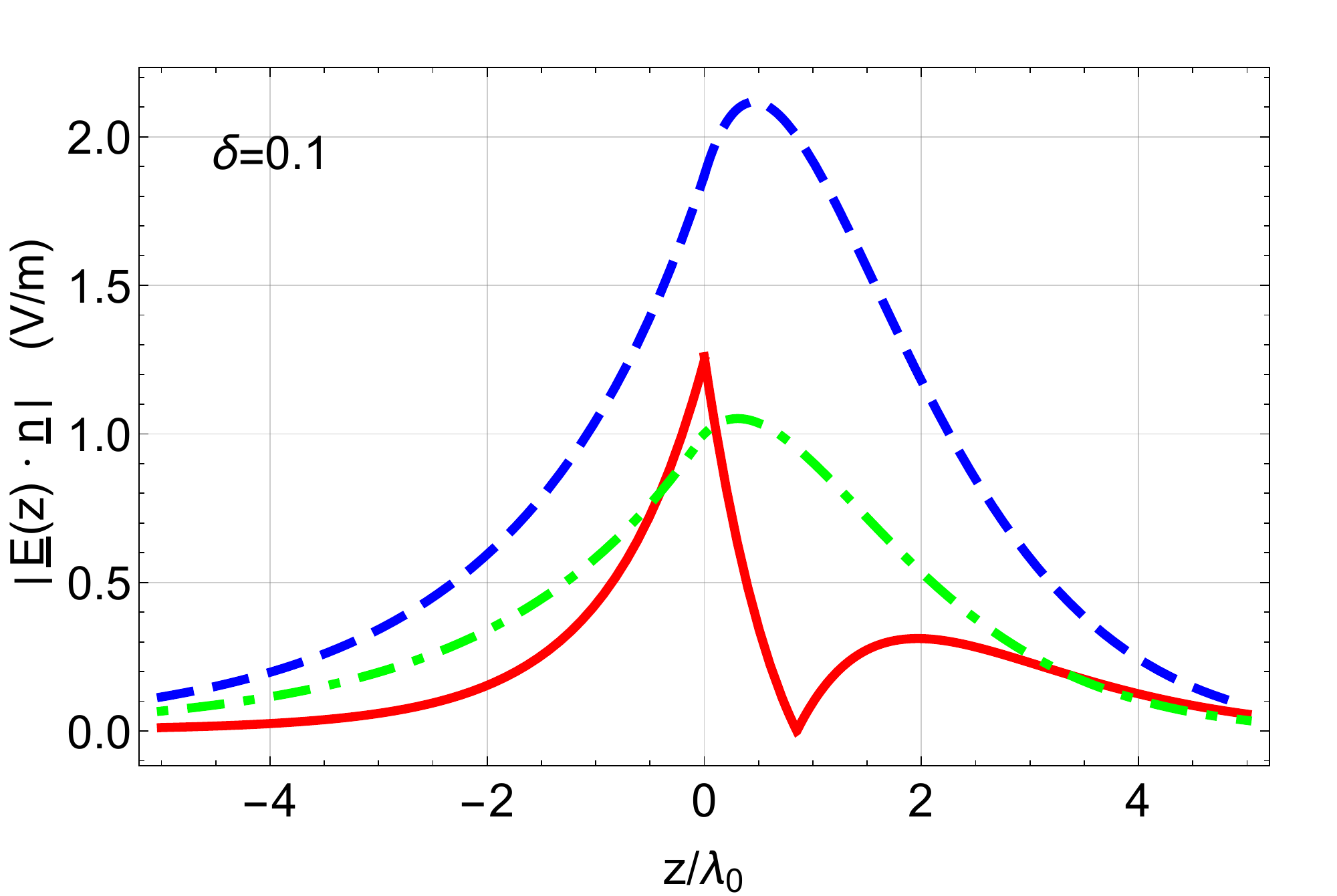}\\
\includegraphics[width=4.3cm]{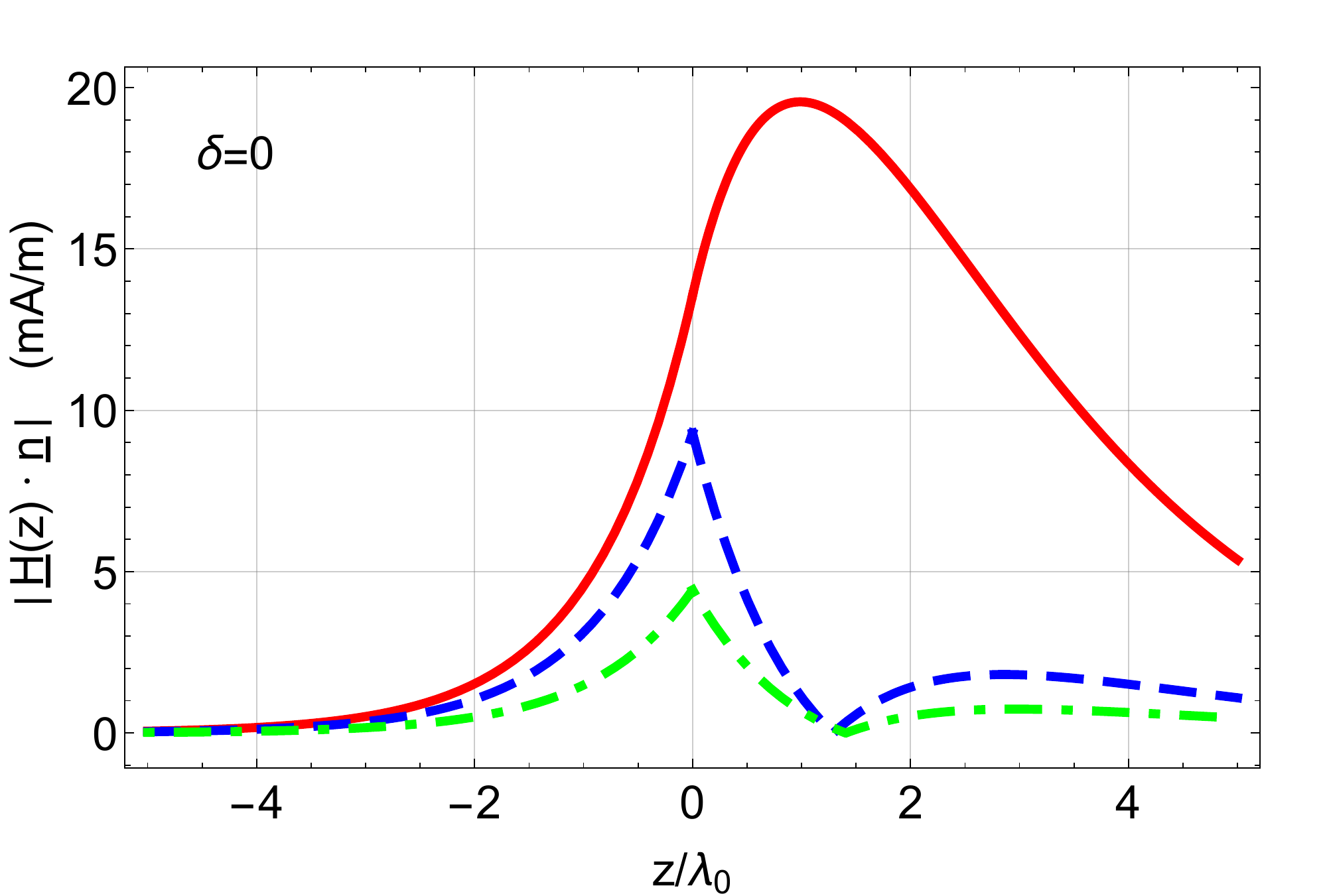}
\includegraphics[width=4.3cm]{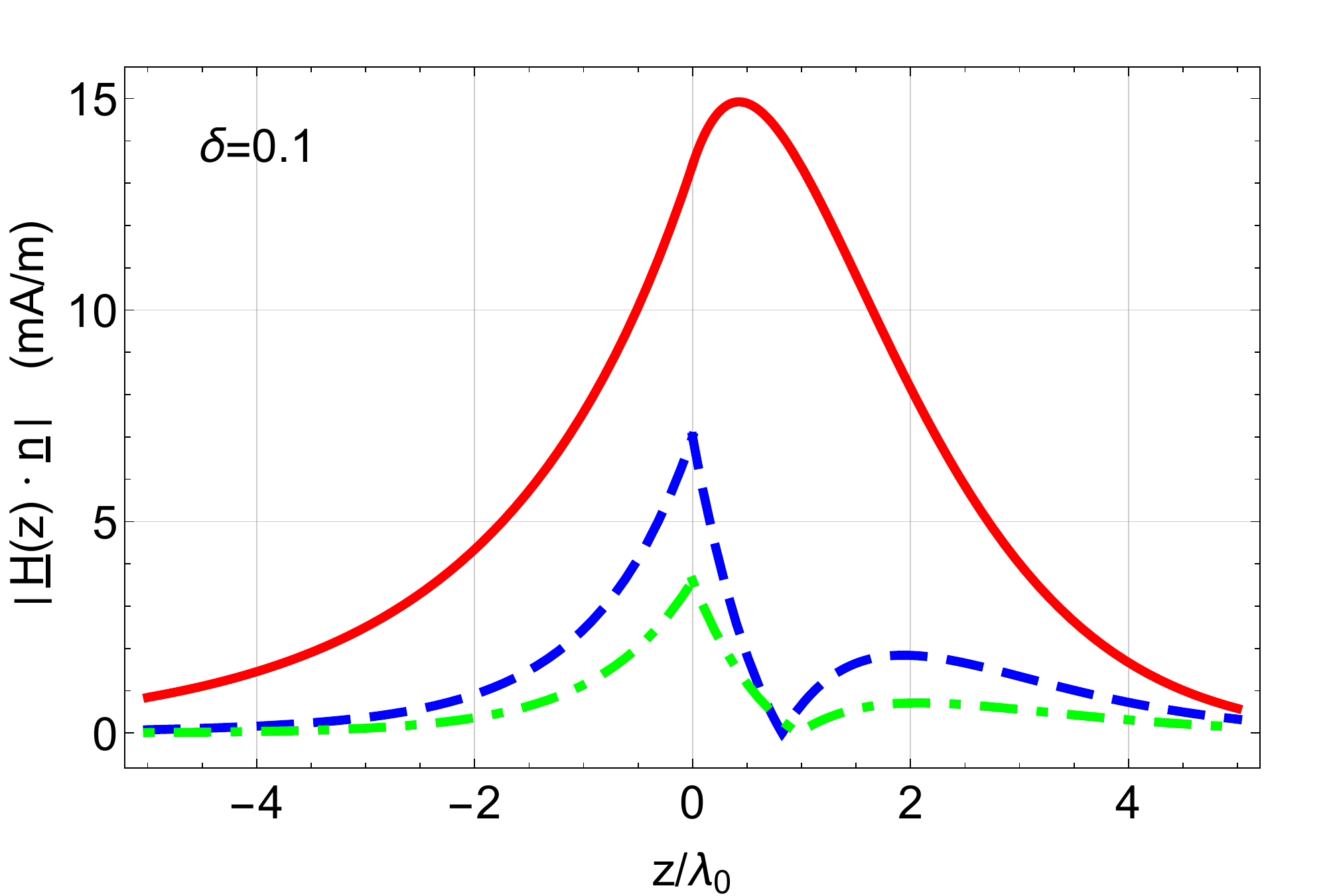}\\
\includegraphics[width=4.3cm]{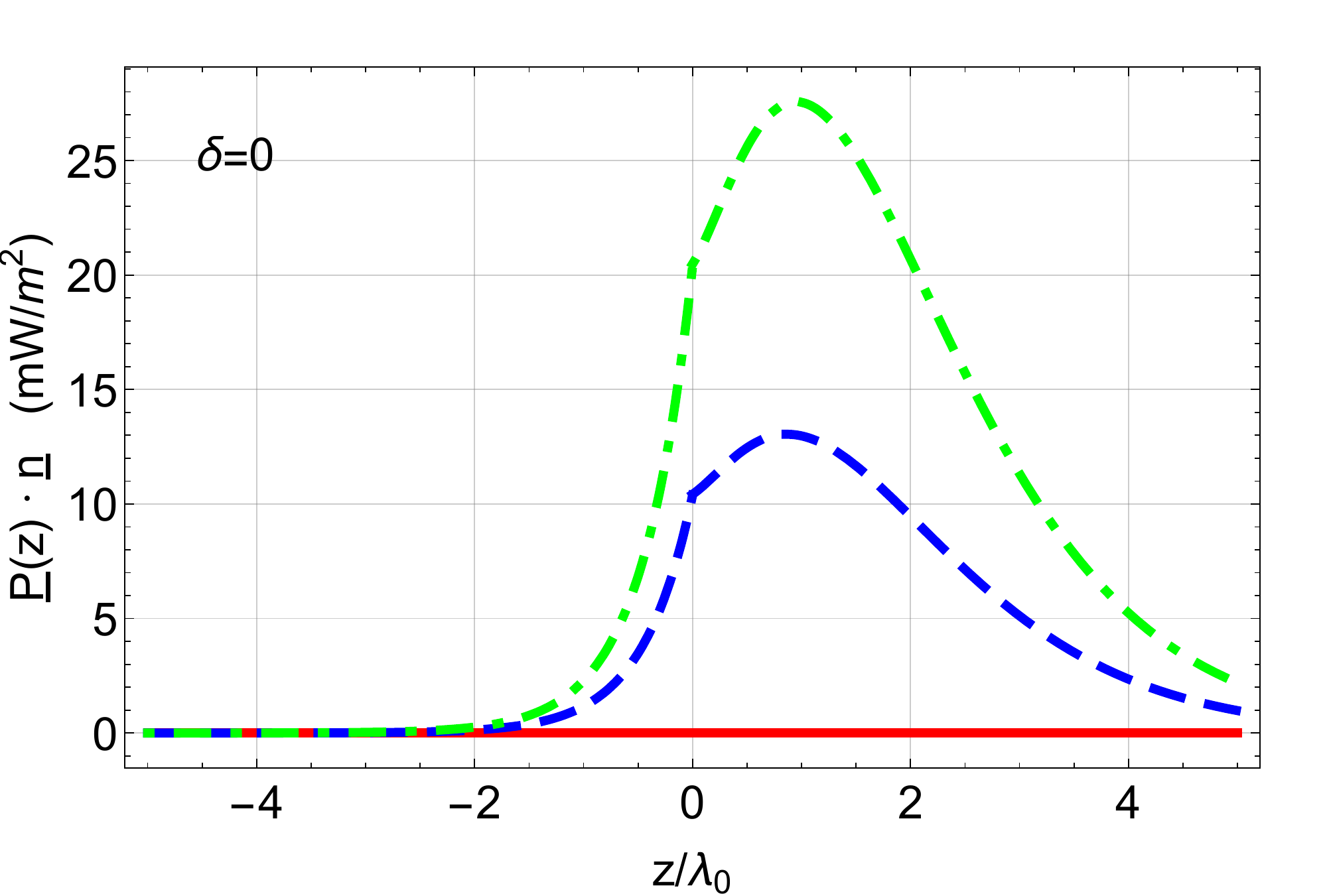}
\includegraphics[width=4.3cm]{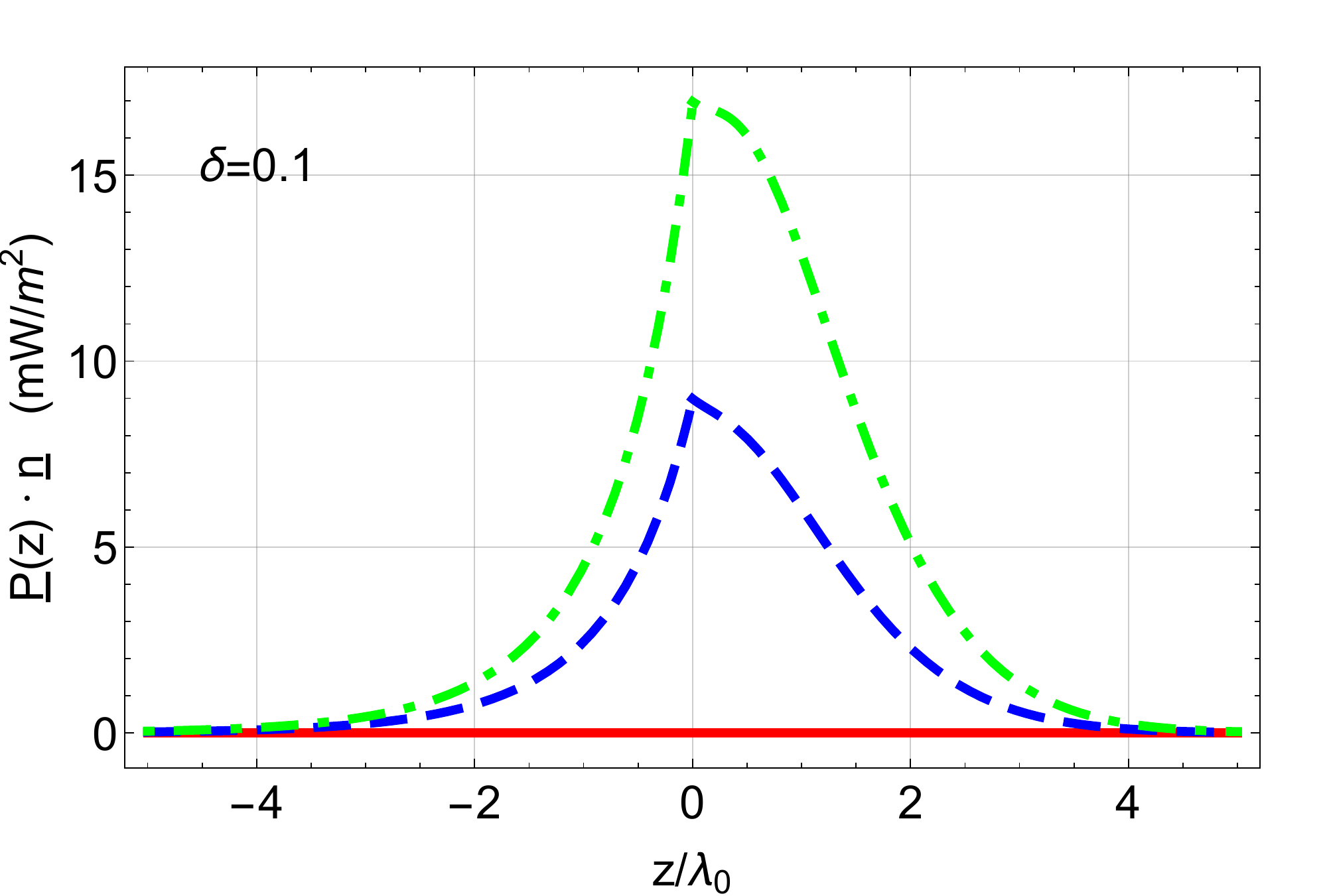}\\
 \caption{
$ | \#E (z \uz) \. \#n |$, $| \#H (z\uz) \. \#n |$, and 
$\#P (z\uz) \. \#n$  plotted versus $z/\lambdao$, 
 when medium $\calA$ is crocoite and medium $\calB$ is specified by $\epsBa = 5.6791$ and $\epsBb = \epsBa + \delta$.
 The propagation angle $\psi = 31^\circ$.
(left) $\delta = 0$ for which $q =2.3894 \ko$; (right) $\delta= 0.1$ for which $q =2.3901 \ko$.
 Normalization is such that $ | \#E ( \#0 ) \. \ux | = 1$~V~m${}^{-1}$.
 Key: $\#n = \ux$ 
 (green broken dashed curves),  $\#n = \uy$ (blue dashed curves), and $\#n = \uz$
 (red solid curves). 
 } \label{fig2}
\end{figure}

Figure \ref{fig2} shows the magnitudes of the Cartesian components of the electric field phasor $\#E(\#r)$, magnetic field phasor
$\#H(\#r)$,
and the time-averaged Poynting vector 
\begin{equation}
\#P(\#r)=(1/2){\rm Re}\lec \#E(\#r)\times \#H^\ast(\#r)\ric
\end{equation}
on the $z$ axis
for  
\begin{itemize}
\item[(i)] the exceptional surface wave existing at $\delta = 0$ and
\item[(ii)] the unexceptional surface wave existing at $\delta = 0.1$.
\end{itemize}
There are marked differences in the localizations of these two surface waves:
The exceptional surface wave is more tightly localized to the interface than the unexceptional surface wave in medium
$\calB$ (i.e., $z<0$),
whereas 
the unexceptional surface wave is more tightly localized to the interface than the exceptional surface wave in
medium $\calA$ (i.e.,  $z>0$).
The maximum energy density flow for the unexceptional surface wave occurs at $z = 0$ while the  maximum
for the exceptional surface wave occurs in the vicinity of $z \approx 0.9 \lambdao$ (in medium $\calA$).

In order  to consider an  exceptional surface wave conforming to Case~III 
next, we   
set $\epsAa=5.6$, $\epsAb=\epsAa + \delta$,  $\epsAc=\epsAa - \delta$,
$\epsBa = 5$, and $\epsBb = 17$. As previously, $\delta \ge 0$. 
The propagation angle $\psi$ is fixed at $19.1183^\circ$.

\begin{figure}[!htb]
\centering
\includegraphics[width=6.3cm]{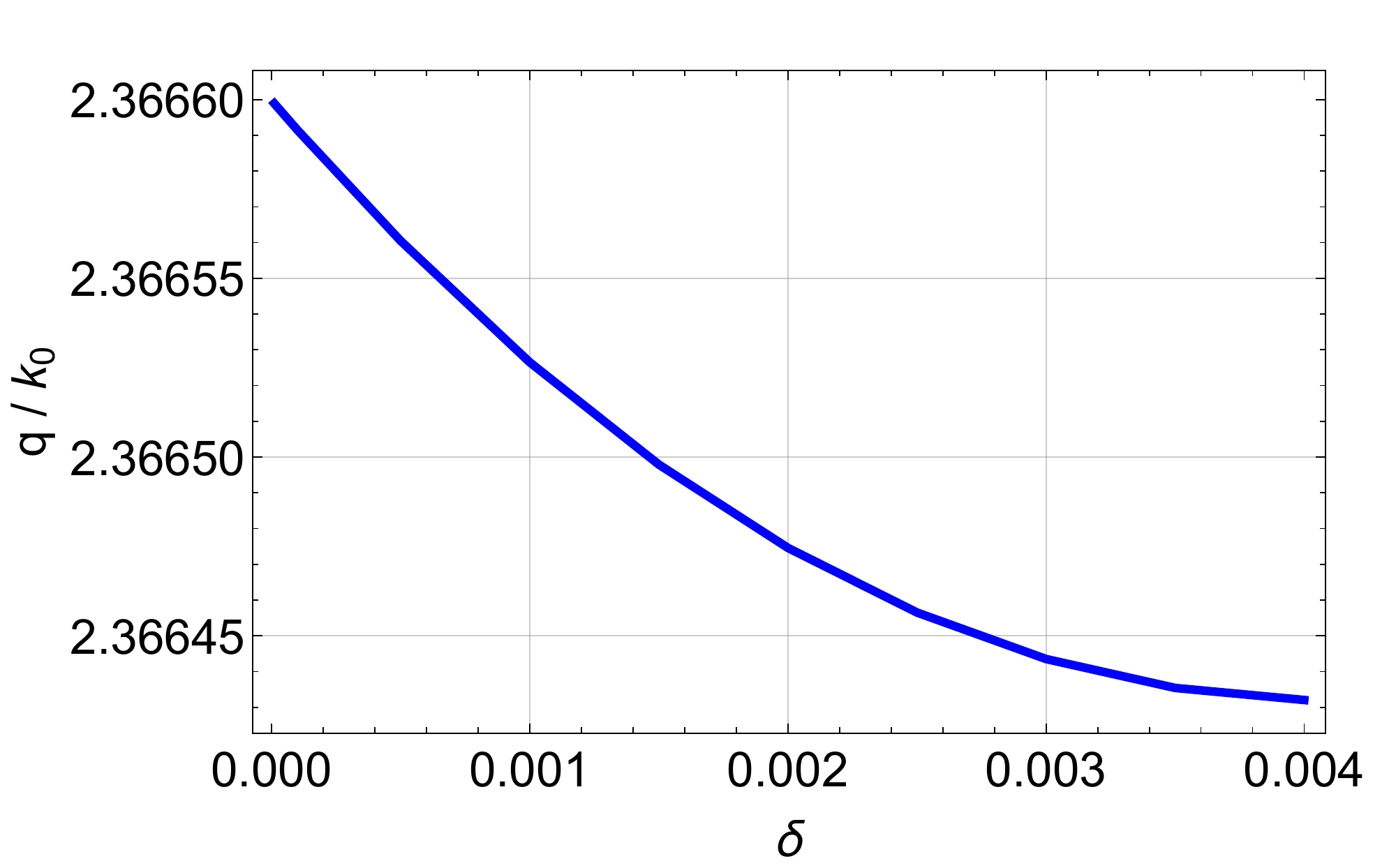}
 \caption{$q/\ko$ as a function of $\delta$ when medium $\calA$ is 
 specified by $\epsAa = 5.6$, $\epsAb = \epsAa + \delta$, and $\epsAc = \epsAa - \delta$, and
   medium $\calB$ is specified by $\epsBa = 5$ and $\epsBb = 17$. The propagation angle 
   $\psi = 19.1183^\circ$.} \label{fig3}
\end{figure}

For this example, 
unexceptional surface waves conforming to Case~I exist for $\delta > 0$ while an
exceptional surface wave conforming to  Case~III exists for $\delta =0$.
For the exceptional surface wave,
the fields in the   half-spaces $z>0$ and $z<0$ are specified through  Eqs.~\r{D_gen_solA} and \r{DV_gen_solB}, respectively.
Again, Eq.~\r{dispersion_eq} does not have to be solved to determine $q$ for the  exceptional surface wave; instead
 $q$ is provided by
Eq.~\r{qb_sol}. Thereafter, Eq.~\r{eq36} has to be manipulated to determine
three of the four coefficients in the column 4-vector on the left side 
in terms of the fourth coefficient.

In Fig.~\ref{fig3}, the graph of
the relative surface wave number $q/\ko$ versus  $\delta$ is displayed  for $\delta \in \les 0, 0.004 \ris$. 
The value of  $q/\ko$ decreases uniformly as $\delta$ increases. No surface-wave solutions were found
for $\delta>0.004$.

\begin{figure}[!htb]
\centering
\includegraphics[width=4.3cm]{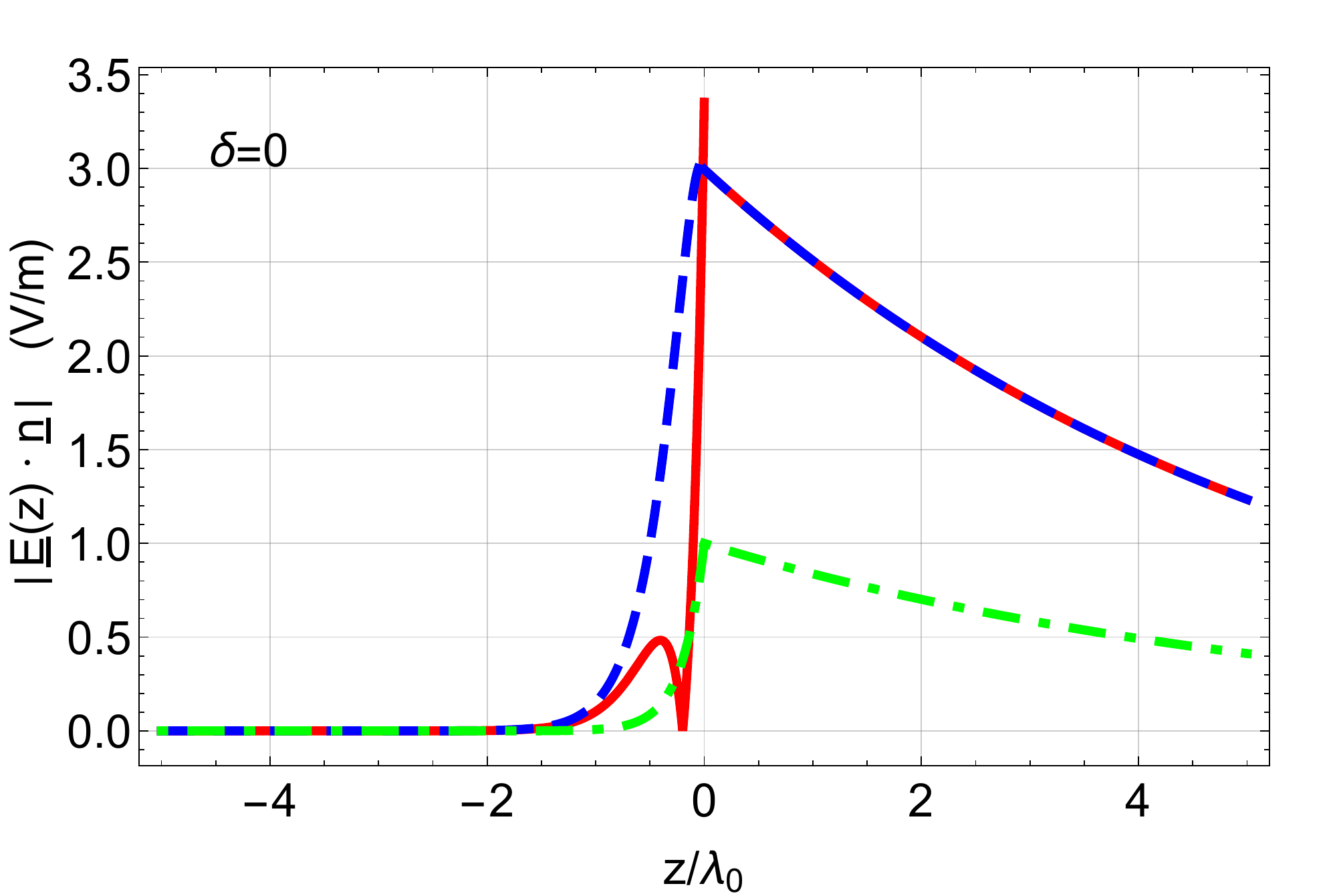}
\includegraphics[width=4.3cm]{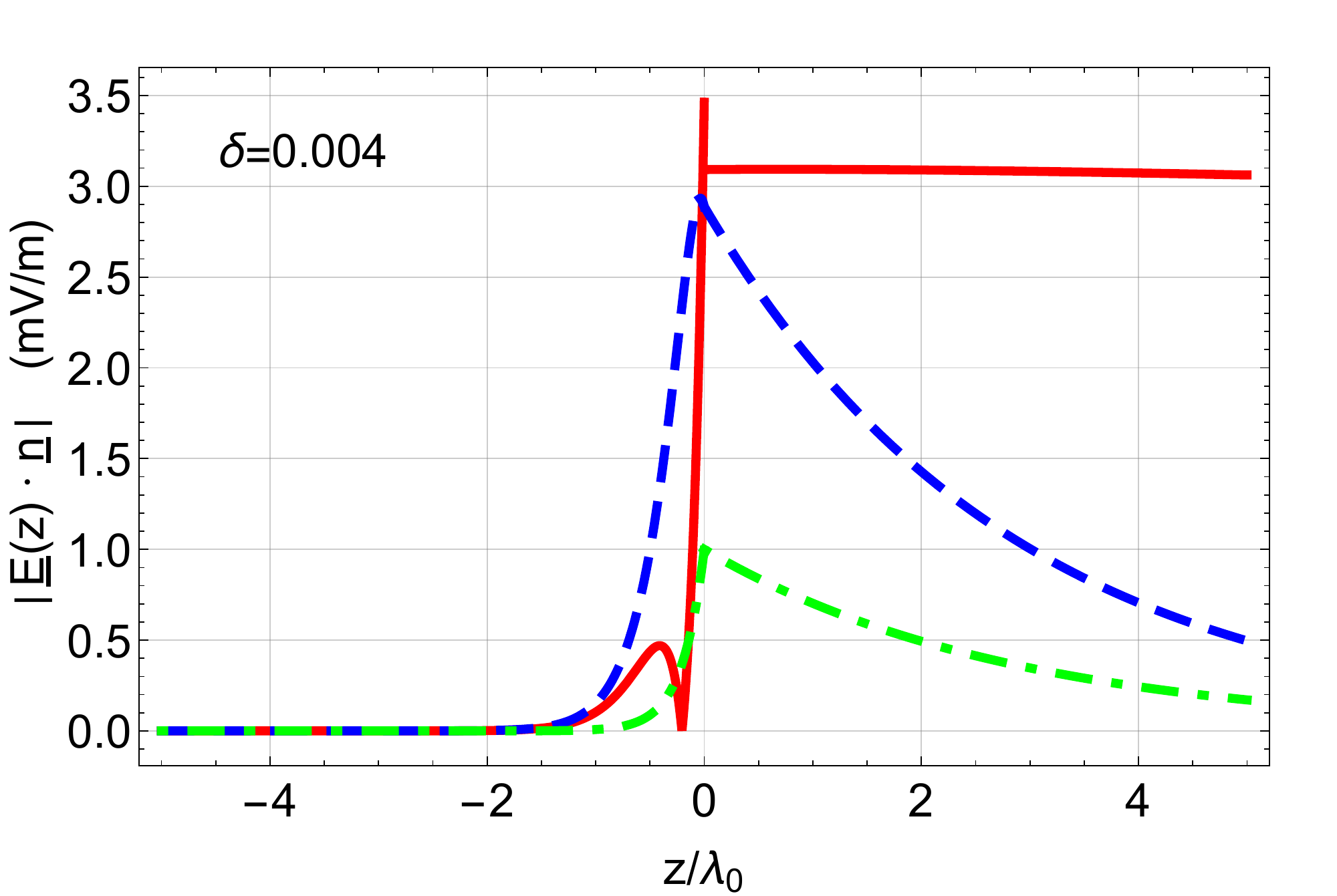}\\
\includegraphics[width=4.3cm]{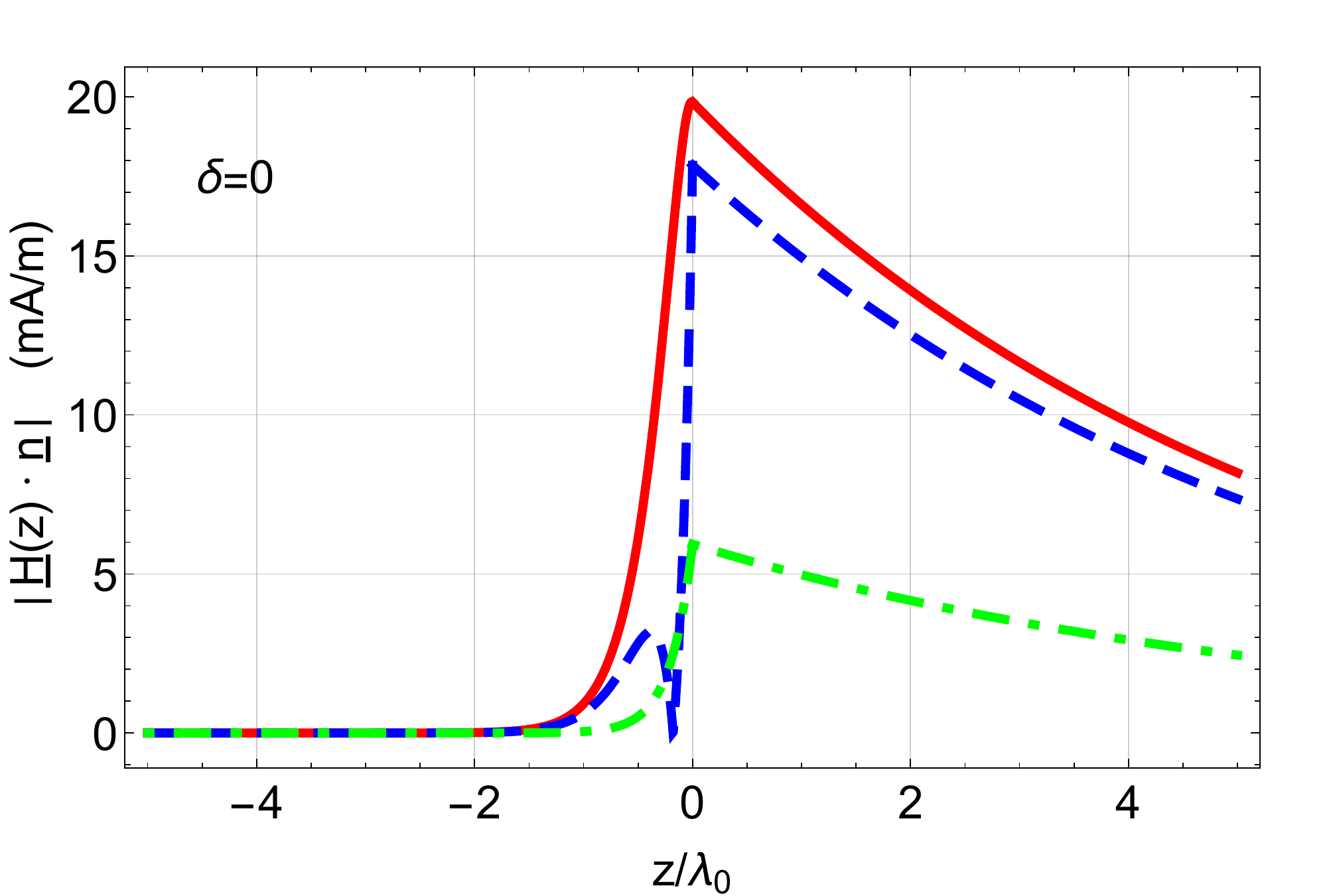}
\includegraphics[width=4.3cm]{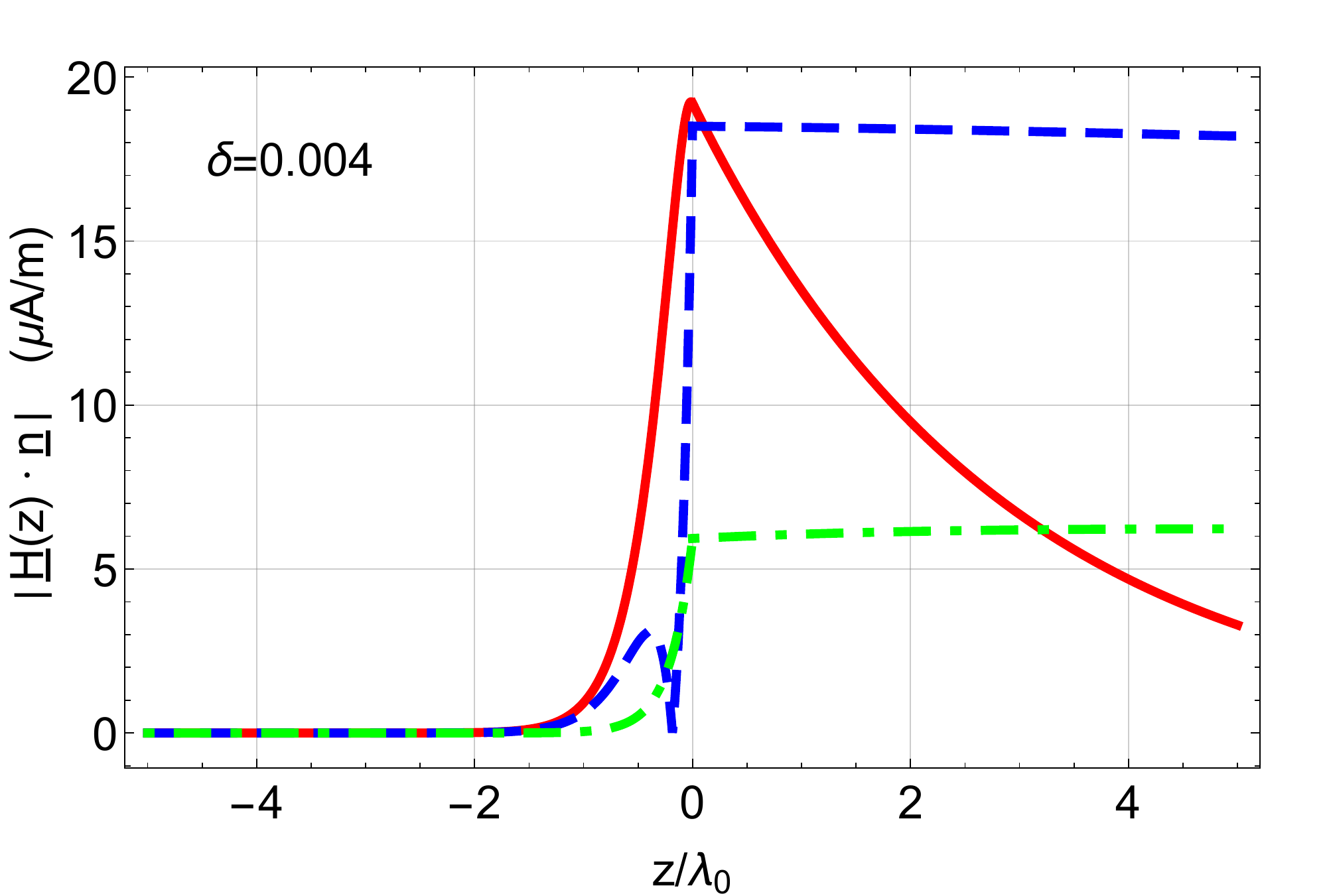}\\
\includegraphics[width=4.3cm]{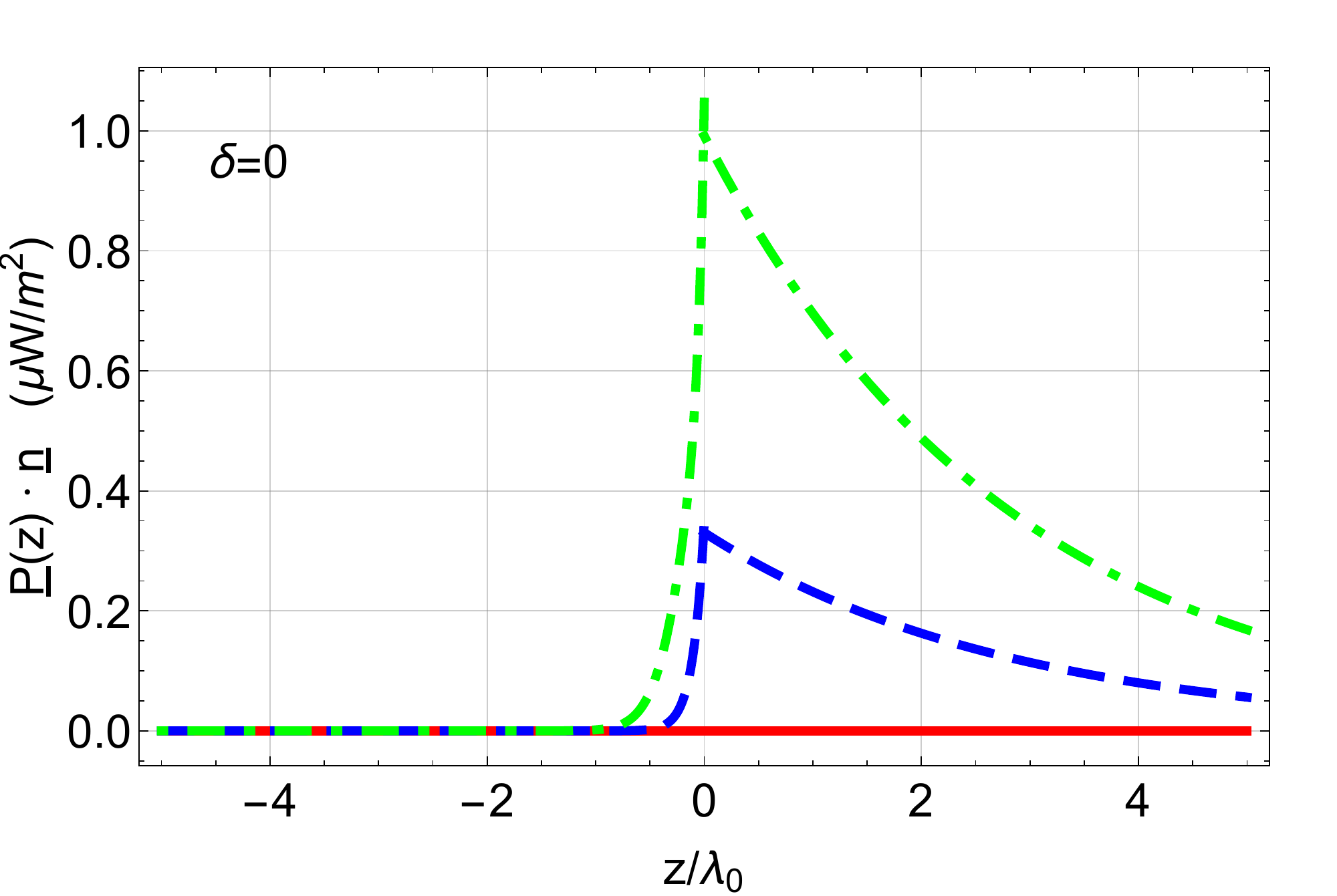}
\includegraphics[width=4.3cm]{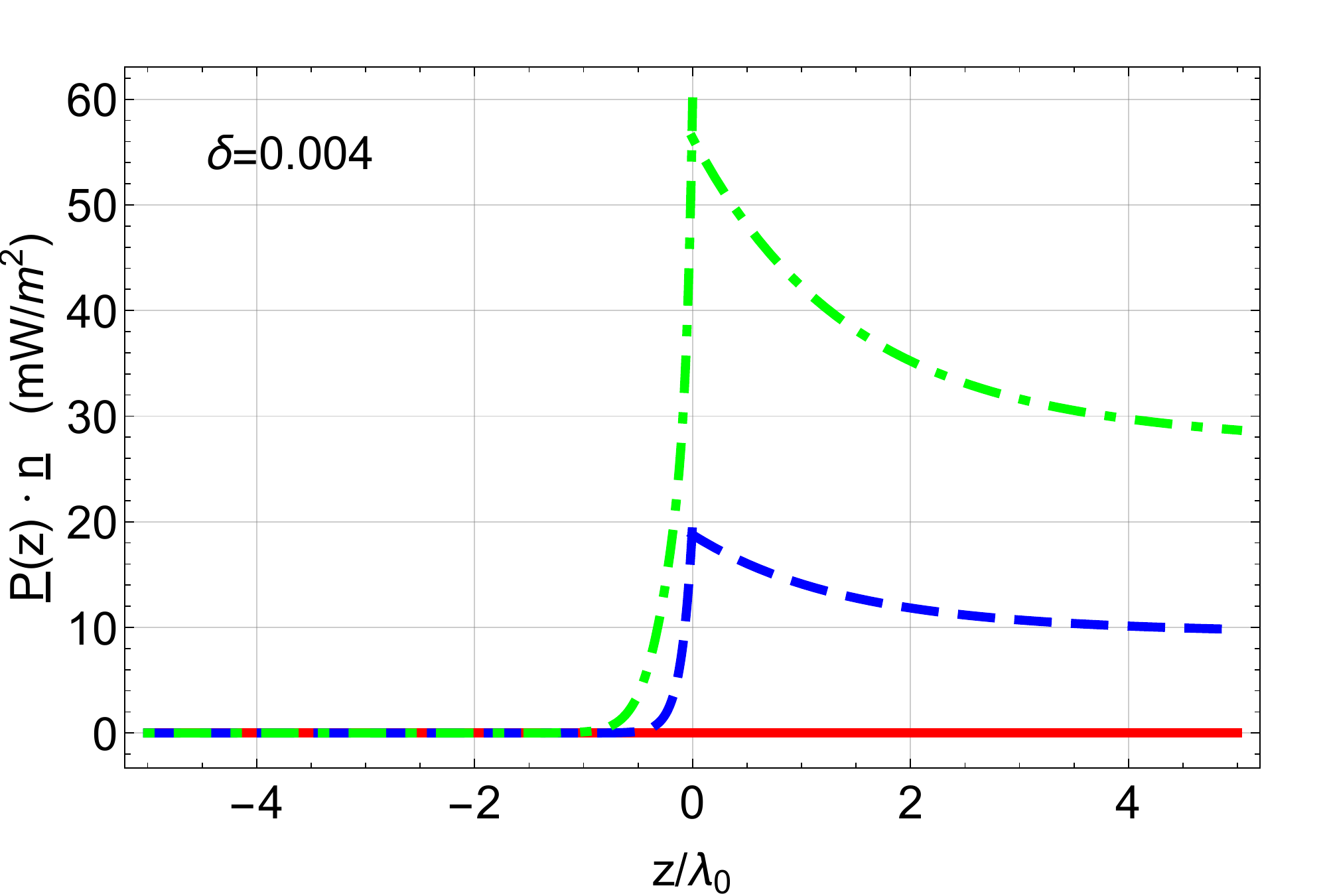}\\
 \caption{
$ | \#E (z \uz) \. \#n |$, $| \#H (z\uz) \. \#n |$, and 
$\#P (z\uz) \. \#n$  plotted versus $z/\lambdao$, 
 when medium $\calA$ 
is  specified by $\epsAa = 5.6$, $\epsAb = \epsAa + \delta$, and $\epsAc = \epsAa - \delta$, 
  and medium $\calB$ is specified by  $\epsBa = 5$ and $\epsBb = 17$.   The propagation angle  $\psi = 19.1183^\circ$.
  (left) $\delta = 0$ for which $q =2.36660 \ko$; (right) $\delta= 0.004$ for which $q =2.36642 \ko$.
  Normalization is such that $ | \#E (z \uz) \. \ux | = 1$~V~m${}^{-1}$.
 Key: $\#n = \ux$ 
 (green broken dashed curves),  $\#n = \uy$ (blue dashed curves), and $\#n = \uz$
 (red solid curves).
 } \label{fig4}
\end{figure}

Profiles of $\#E(z\uz)$,  
$\#H(z\uz)$, and $\#P(z\uz)$
are presented in Fig.~\ref{fig4} for
\begin{itemize}
\item[(i)] the exceptional surface wave existing at $\delta = 0$ and
\item[(ii)] the unexceptional surface wave existing at $\delta = 0.004$.
\end{itemize}
 The profiles for the two surface waves are qualitatively similar but differences in the degree of localization are apparent:
The exceptional surface wave is more tightly localized to the interface than the unexceptional surface wave in medium
$\calA$  (i.e., $z>0$),
whereas the degree of localization in medium $\calB$ (i.e., $z<0$) is approximately the same for the 
 unexceptional and exceptional surface waves.   

\begin{figure}[!htb]
\centering
\includegraphics[width=4.3cm]{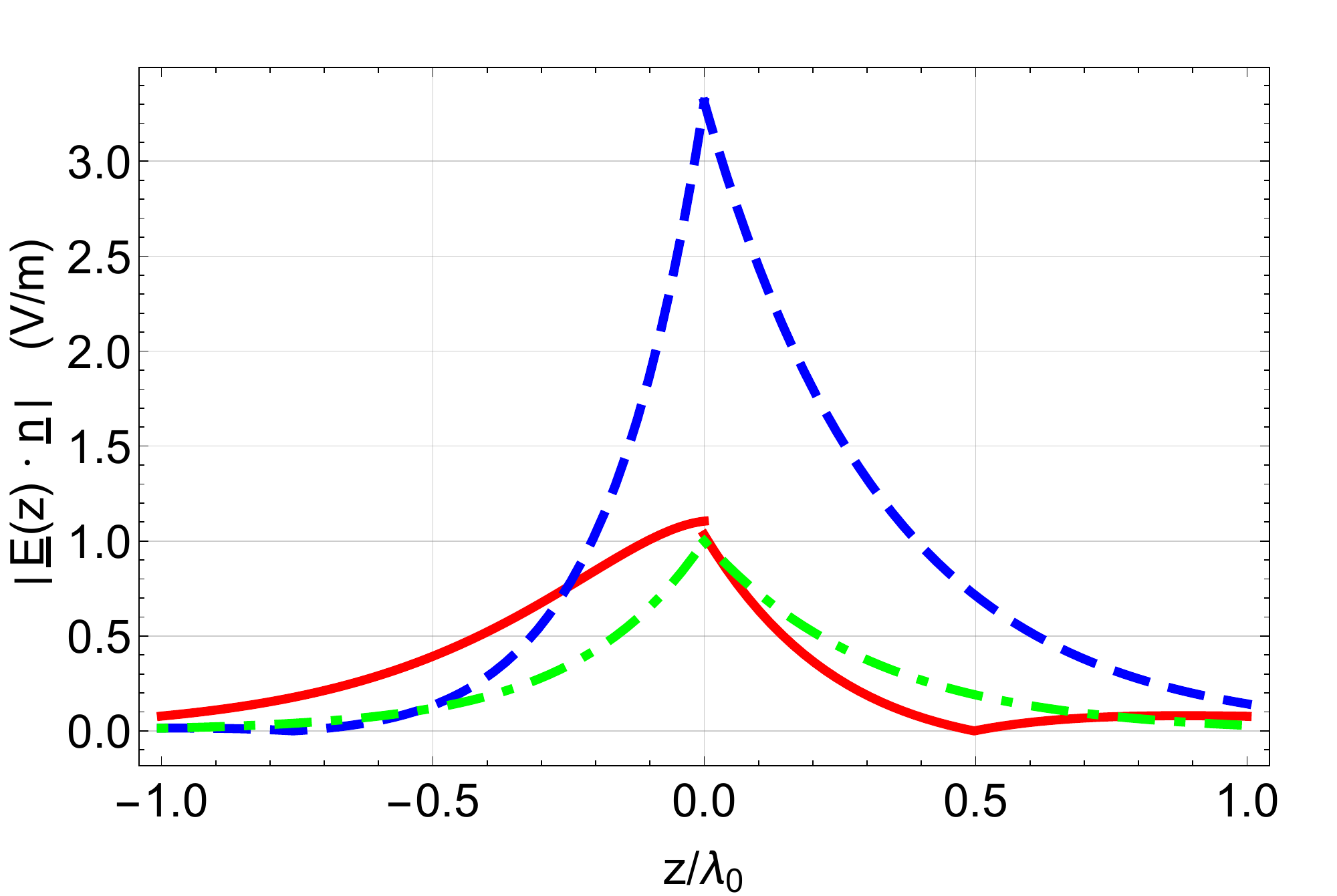}
\includegraphics[width=4.3cm]{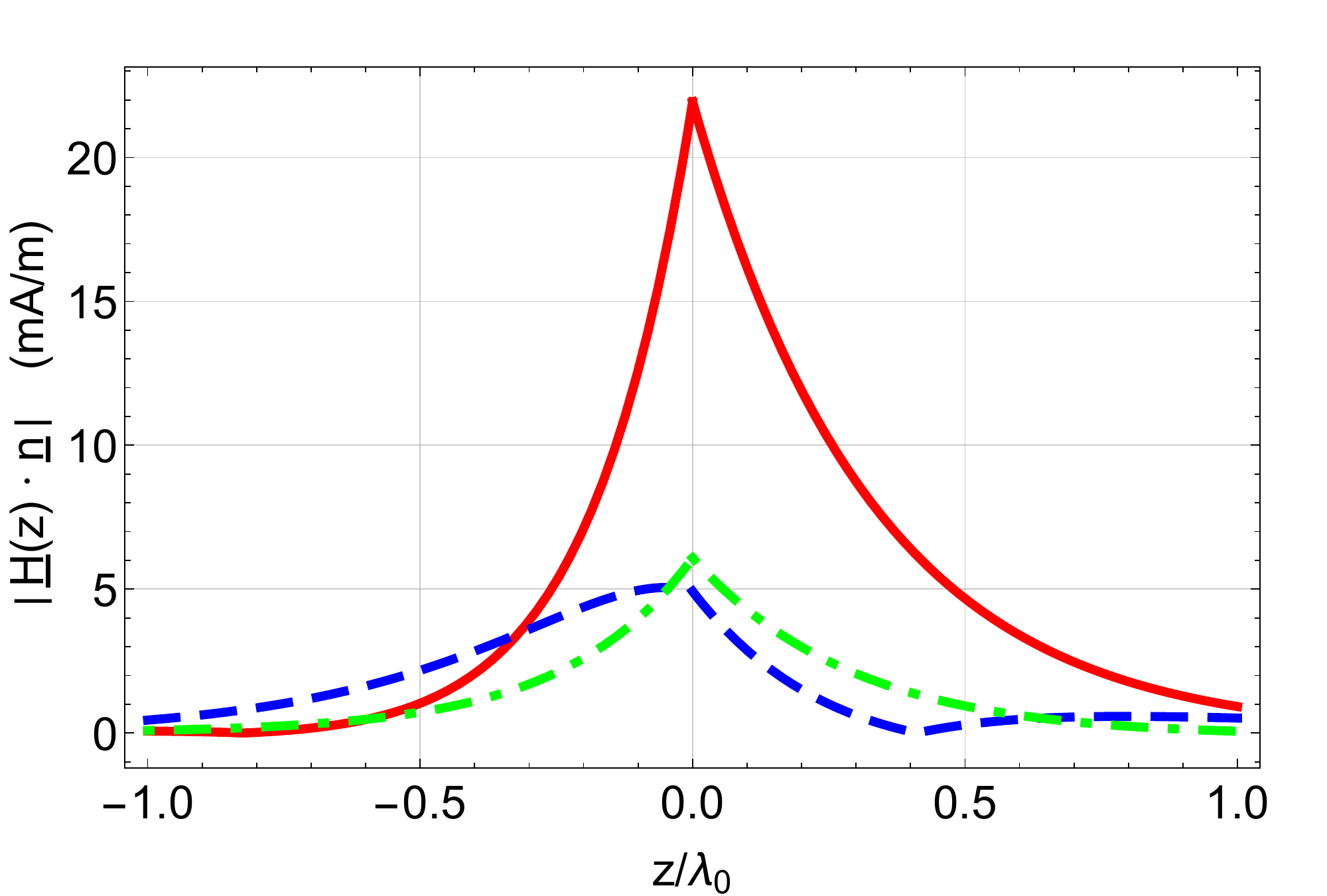}\\
\includegraphics[width=4.3cm]{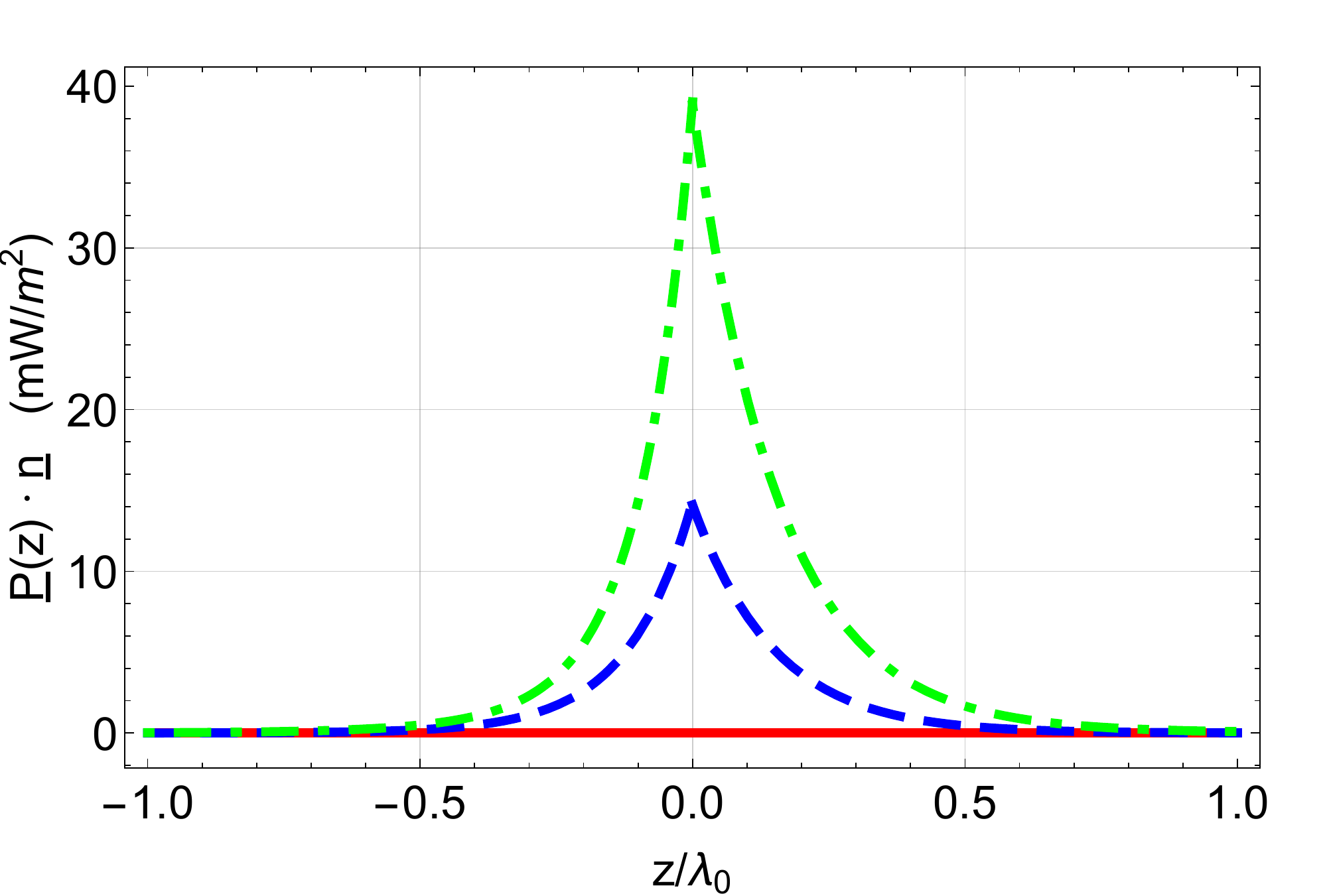} \caption{
$ | \#E (z \uz) \. \#n |$, $| \#H (z\uz) \. \#n |$, and 
$\#P (z\uz) \. \#n$  plotted versus $z/\lambdao$, 
 when  medium $\calA$ is crocoite and medium $\calB$ is specified by  $\epsBa = 5.25$ and $\epsBb = 8.4379$. 
 The propagation angle  $\psi = 16.4480^\circ$ and the relative wavenumber $q =2.3891 \ko$.
 Normalization is such that $ | \#E (z \uz) \. \ux | = 1$~V~m${}^{-1}$.
 Key: $\#n = \ux$ 
 (green broken dashed curves),  $\#n = \uy$ (blue dashed curves), and $\#n = \uz$
 (red solid curves).
 } \label{fig5}
\end{figure}

Lastly, we turn to Case~IV, i.e.,  the case of a doubly exceptional surface wave. For this purpose, we again take medium $\calA$ to be
 crocoite in its orthorhombic form while medium $\calB$ is specified by the relative permittivity parameters  $\epsBa = 5.2500$ and $\epsBb = 8.4379$.
 The propagation angle  $\psi = 16.4480^\circ$.
In this case  the surface wavenumber $q$
must simultaneously satisfy Eqs.~\r{qa_sol} and \r{qb_sol};
its value is $q = 2.3891 \ko$.
The spatial profiles of the field phasors available from
from Eq.~\r{eq36} are plotted in Fig.~\ref{fig5}.
These spatial profiles are qualitatively and quantitatively different from those for the exceptional surface waves in Figs.~\ref{fig2} and \ref{fig4}.
Specifically, the doubly exceptional surface wave is very tightly bound to  the interface, being largely
confined to the   region $\vert{z}\vert < \lambdao$.
 Also, the energy density flow for the doubly exceptional surface wave is approximately symmetric 
across the bimedium interface $z=0$,
 in contrast to that for  
the exceptional surface waves  in  Figs.~\ref{fig2} and \ref{fig4}.

Unlike the exceptional surface waves (Cases~II and III), the doubly exceptional surface wave  arises as an 
isolated degeneracy in the space of the constitutive parameters and propagation angle. That is,  if the 
constitutive parameters or the propagation angle 
that yield a  doubly exceptional surface wave
are varied slightly, even by a minuscule amount (say $< 0.0001 \%$), then no surface-wave solutions can be found.

\section{Concluding Remarks}\label{cr}

The planar interface of two dissimilar homogeneous dielectric  mediums can guide exceptional surface waves when one of the
two partnering mediums is anisotropic while the other is isotropic (but we note that other constitutive
contrasts between the two partnering mediums may also allow exceptional surface waves to exist).
Exceptional surface waves are distinguished from unexceptional surface waves by their unique localization characteristics in the anisotropic partnering medium.
If both partnering mediums are anisotropic, we have shown here that a doubly exceptional surface wave could exist for an isolated propagation direction.
Doubly exceptional surface waves are distinguished from unexceptional surface waves 
and exceptional surface waves
by their unique localization characteristics in both partnering mediums.

The  numerical demonstration of a 
doubly exceptional surface wave 
presented herein
was based on anisotropic partnering mediums characterized by physically realizable constitutive parameters.
Furthermore, our numerical studies revealed that the doubly exceptional surface wave  arises as an isolated degeneracy in the space of the constitutive parameters and the propagation direction; i.e., the  doubly exceptional surface wave is 
enclosed by null surface-wave solutions in the space of the constitutive parameters and the propagation direction.

\vspace{5mm}

\noindent {\bf Acknowledgments.}
This work was supported in part by
EPSRC (grant number EP/S00033X/1) and US NSF (grant number DMS-1619901).
AL thanks the Charles Godfrey Binder Endowment at the Pennsylvania State University    for partial support of his research endeavors.

\vspace{5mm}

\noindent {\bf Disclosures.} The authors declare no conflicts of interest.

\end{document}